\documentclass[a4paper,12pt]{article}
\usepackage{amsmath,amsthm,amsfonts,amssymb,bm,mathrsfs}
\usepackage{enumerate}
\usepackage{graphicx}
\usepackage{natbib}
\usepackage{xcolor}
\usepackage{sgame}
\usepackage[text={7in,10in},centering]{geometry}
\usepackage{tikz}
\usepackage{float}
\usepackage{setspace}
\usepackage{hyperref}
\usepackage{comment}
\usepackage{comment}

\usetikzlibrary{patterns}
\renewcommand{\baselinestretch}{1.35}
\allowdisplaybreaks
\newtheorem{assum}{Assumption}

\newtheorem{lem}{Lemma}
\newtheorem{prop}{Proposition}
\newtheorem{rmk}{Remark}
\newtheorem{exam}{Example}

\hypersetup{colorlinks=true,
citecolor=blue,
linkcolor=blue,
urlcolor=blue,
bookmarksopen=true,
pdfstartview={XYZ null null 1.00},
pdfpagelayout={SinglePage}
}
\DeclareMathOperator{\rmd}{d\!}

\DeclareFixedFont{\myfont}{OT1}{pnc}{m}{n}{14pt}

\usepackage{hyperref}

\hypersetup{colorlinks=true,
            citecolor=blue,
            linkcolor=blue,
            urlcolor=blue,
            bookmarksopen=true,
            pdfstartview={XYZ null null 1.00},
            pdfpagelayout={SinglePage}
}

\begin{document}

\title{Learning by Consuming: Optimal Pricing with Endogenous Information Provision\thanks{We are grateful for helpful discussions with Tilman B\"orgers, Ilwoo Hwang, Silvana Krasteva, R. Vijay Krishna, Rohit Lamba, Jingfeng Lu, Sergei Severinov, Vasiliki Skreta, Roland Strausz, Siyang Xiong, Jidong Zhou, as well as audiences at TAMU, UC Riverside, Wuhan University, Sun Yat-sen University, 2022 Decentralization Conference, 2022 North American Summer Meeting of the Econometric Society, 2022 Conference on Mechanism and Institution Design, and 21st SAET conference. He and Liu acknowledge financial support from the Research Grants Council of Hong Kong (14503318) and the National Natural Science Foundation of China (72192805 and 72073115), respectively.
}}
\date{\today}
\author{Huiyi Guo\thanks{%
Huiyi Guo: Department of Economics, Texas A\&M University, 4228 TAMU,
College Station, TX 77843, United States of America. Email: 
\mbox
{huiyiguo@tamu.edu.}} \and Wei He\thanks{%
Wei He: Department of Economics, The Chinese University of Hong Kong, Hong
Kong, China. Email: \mbox
{hewei@cuhk.edu.hk.}} \and Bin Liu\thanks{%
Bin Liu: School of Management and Economics and Shenzhen Finance Institute,
The Chinese University of Hong Kong, Shenzhen (CUHK-Shenzhen), China 518172.
Email: \mbox
{binliu@cuhk.edu.cn.} } 
}

\maketitle

\begin{abstract}
\thispagestyle{empty}%
\def\baselinestretch{1.1}\footnotesize%
We study the revenue-maximizing mechanism when a buyer's value evolves endogenously because of learning-by-consuming. A seller sells one unit of a divisible good, while the buyer relies on his private, rough valuation to choose his first-stage consumption level. Consuming more leads to a more precise valuation estimate, after which the buyer determines the second-stage consumption level. The optimum is a menu of try-and-decide contracts, consisting of a first-stage price-quantity pair and a second-stage per-unit price for the remaining quantity. In equilibrium, a higher first-stage valuation buyer pays more for higher first-stage consumption and enjoys a lower second-stage per-unit price. Methodologically, we deal with the difficulty that due to the failure of single-crossing condition, monotonicity in allocation plus the envelope condition is insufficient for incentive compatibility. Our results help to understand contracts about sequential consumption with the learning feature; \textit{e.g.}, leasing contracts for experience goods and trial sessions for certain courses.

\end{abstract}



\newpage

%


\section{Introduction}

Situations are abundant in which a consumer is uncertain about how well the good's characteristics fit him at the outset, but by consuming (a portion of) the good, he could learn additional information to refine the value estimation. With such a more precise value estimate, the consumer then decides how many further units of the good to consume. Such kind of \textit{learning by consuming} is widely observed in practice.

For example, car dealers in the U.S. usually provide a menu of contracts to potential consumers. Some consumers may choose to enter leasing contracts, which give them both the right to drive the car during the lease term and a lease-end option to buy out the car. The cost of the leasing contract and the buyout price depend on the length of the lease term, which usually ranges from two to four years. These consumers can learn their matching values with the car in the lease term, before deciding whether to buy out the car or not. Consumers who are more uncertain are often more willing to learn matching values by entering the leasing contracts. Other consumers may choose to buy the car outright instead. This usually happens when a consumer is sufficiently optimistic about the matching value, and hence wishes to secure a lower payment for the long-term ownership of the car.

As another example, when purchasing certain courses with fixed terms --- a one-month package of fitness classes from a gym, a two-month playgroup for pre-school toddlers, or a summer sports course for children --- the consumer often prefers to experience a few sessions first. After paying a fee and attending a few included sessions which can be viewed as trial sessions, the consumer refines his valuation and decides whether to register for additional (or the remaining) sessions or not. As in the car-leasing example, in practice, the seller often sets a price for the included sessions and another price for additional sessions, and typically both prices depend on the number/length of included sessions. It is commonly observed that the seller offers different pricing packages to the consumer, who then decides which package to choose.\footnote{For example, Orangetheory Fitness, a popular fitness chain with more than a million members in the U.S., offers three monthly membership packages for consumers. These options differ with each other mainly in terms of the number of included sessions and fee structures.}


There are two important features in the examples above. First, the seller can choose to first sell a portion of the good to the consumer, through which the consumer better understands how well the good fits him and then decides the subsequent consumption. Second, since the consumer's learning is achieved through consuming, there is a tension between information acquisition and future consumption. Naturally, consuming more in the beginning would lead to a more precise value estimate, so that he could make a better decision in the future. However, at the same time, it also means that the size of the remaining portion of the good decreases. For instance, while entering a contract with a long lease term induces sufficient learning, it may make the buy-out option unattractive as the car will be getting old; experiencing more trial sessions helps the consumer better learn the valuation but decreases the number of sessions that can be sold beyond the trial.


How should the seller incorporate such kind of learning by consuming into her selling mechanism? To address this question, we study a two-stage model, in which a risk-neutral seller sells one unit of a \textit{divisible} good to a risk-neutral buyer. The buyer's valuation depends on how well
the good fits him, which is uncertain to him at stage one. Yet, at stage one, he has a prior -- rough private valuation of the good. Relying on this rough valuation, he decides how many units to experience. Experiencing the good provides him with additional private information regarding the good's characteristics. Consuming more leads to more precise additional information.\footnote{The precision of the additional information through consumption is defined in the sense of rotation order, see Section~\ref{sec:model} for details. } With the updated private valuation, the buyer further determines his second-stage consumption level.

In our problem, the first-stage allocation (\textit{i.e.}, consumption)\footnote{We use ``allocation'' and ``consumption'' interchangeably.} plays two roles. First, it is the device for information acquisition, since it provides the buyer with additional information, which will be more precise with a higher first-stage allocation. Meanwhile, it also defines an \emph{intertemporal} problem: It ``secures" some consumption in the early period, regardless of whether the newly acquired information is good or bad, and also determines the maximum amount of consumption in the later stage. Clearly, a
revenue-maximizing seller should incorporate both roles of the first-stage allocation into her pricing strategy.

We fully characterize the revenue-maximizing mechanism and find that the
optimum can be implemented by a menu of \emph{try-and-decide}  contracts, consisting of a
first-stage price-quantity pair and a second-stage per-unit price for the
remaining quantity. When the buyer selects some try-and-decide contract, he needs to pay the corresponding first-stage price specified by the contract chosen. By doing this, the buyer not only gets to experience the corresponding first-stage portion (\textit{i.e.}, quantity) of the good, but also obtains the option to buy the remaining portion at the prescribed second-stage per-unit price. In the optimal contract, a larger first-stage consumption level (quantity) is paired
with a higher first-stage price but a lower per-unit second-stage price for
the remaining portion of the good. Moreover, if the buyer ends up buying
the entire good across two stages, a higher first-stage consumption leads to a lower total payment. In equilibrium, a higher first-stage valuation buyer pays more to consume more in the first stage, in exchange for a lower per-unit price for the remaining portion.

The intuition is as follows. For a high first-stage valuation buyer, he is
more confident that his updated valuation of the good is sufficiently high,
so that he will likely buy the entire portion of the good. 
Thus, he is willing to pay to consume more in the first stage to enjoy a
lower price for the additional consumption in the second stage, and also a
lower total price for the entire portion of the good. 
However, this is quite risky for a low first-stage valuation buyer. If he does so, despite the
second-stage per-unit price being lower, he has to pay to consume more in the first
stage to enjoy this second-stage benefit. Yet, since his first-stage
valuation is low, he really wants to experience the good just a bit to improve his decision in the second stage, rather than \textquotedblleft
blindly\textquotedblright\ having a high first-stage consumption level,
which can lead to a low expected payoff given the low first-stage
valuation.

The format of our optimal try-and-decide contracts resembles practical contracts. For instance, in the car-leasing example, the leasing price can be viewed as the first-stage price, while the length of the lease term and the buyout price can be regarded as the first-stage quantity and second-stage price. In the other course registration example, the trial sessions and the remaining sessions can be viewed as the consumption at the two stages, respectively. In this sense, we provide a rationale for the common phenomena of sequential consumption with learning in reality.


On the technical side, we would like to point out that when solving the optimal mechanism, establishing the global incentive compatibility (IC) condition is quite involved in our setting. In canonical sequential screening problems, \textit{e.g.}, \cite{courty2000sequential} and \cite{esHo2007optimal}, building on local IC, the monotonicity of an allocation rule leads to global IC, even if it may not be the optimal allocation rule. By imposing certain regularity conditions, one can check that the optimal allocation rule in the relaxed problem, which only uses local IC, is indeed monotone. As a result, such an allocation rule also satisfies global IC, so that it is indeed the optimum. However, this standard approach does not apply in our setting. In fact, we provide an example where the first-stage allocation rule is monotone, but this allocation rule cannot be part of a global IC mechanism.  To establish global IC of our mechanism, we have to explicitly make use of the optimality of the first-stage allocation rule. 
The difficulty we encounter can be better explained by focusing on the first-stage problem, where the sequential screening problem can be understood as a corresponding static screening problem; see  \cite{krahmer2017sequential} for further discussions. In the corresponding static problem, a standard condition in the literature, often called the single-crossing condition (alternatively, the constant sign condition, or the Spence--Mirrlees condition), is missing. The lack of such kind of a condition prevents us from establishing global IC only from the monotonicity of the first-stage allocation rule.\footnote{\cite{krahmer2017sequential} show that the condition of first-order stochastic dominance (FOSD) in the canonical sequential screening problem is equivalent to the usual single-crossing condition in a certain static screening problem. As is well known, without the single-crossing condition, solving the optimal mechanism is generally believed to be challenging even in static screening problems; see, for example, \cite{araujo2010adverse} and \cite{schottmuller2015adverse}. In our setting, the standard approach fails because of the violation of the single-crossing condition. This approach may also fail when the optimal solution to the standard relaxed problem is non-monotone; see \cite{krahmer2015ex}, \cite{battaglini2019optimal}, \cite{lu2021optimal}, and \cite{li2022stochastic} for the analysis of optimal mechanisms in this case.}


Our paper joins the growing literature on dynamic mechanism design.\footnote{This is now an extensive literature; please refer to \cite{bergemann2019dynamic} for an excellent survey.} The
canonical literature typically assumes that the agent has two (or more)
stages of private information, where the distribution of the second-stage
private information is exogenously determined by the first-stage private
information. In particular, it is often assumed that a higher first-stage
type corresponds to a better distribution of the second-stage type in the sense
of FOSD. See, for example, \cite{courty2000sequential}, \cite{esHo2007optimal}, \citet{krahmer2015ex,krahmer2017sequential}, and more recently \cite{li2022stochastic}, as well as \cite{battaglini2005long} and \cite{garrett2012managerial} for infinite stages. 
However, in the current work, the distribution of the
second-stage valuation (type) depends on the first-stage consumption,
which is endogenously chosen by the buyer. Due to rotation ordering,\footnote{Studies that also use this information order includes, for example, \cite{johnson2006simple}, \cite{hoffmann2011pre}, and \cite{shi2012optimal}.} such endogenously generated
second-stage information is no longer ranked in terms of FOSD, which is
a feature that does not exist in many canonical papers.\footnote{In \cite{liu2018pairing}, the second-stage type's distribution is also
endogenous (determined by moral hazard), but it is ranked by FOSD.}

Within the dynamic mechanism design literature, there is a strand that
involves information acquisition and provision. Among these papers, the channel of
information acquisition is typically independent of the product sold by the
principal. One approach to model information acquisition in the literature
assumes that there is an outside source of information acquisition. For
example, in 
\citet{esHo2007optimal,
esHo2007price}, \cite{li2017discriminatory}, and \cite{guo2022optimal}, 
 the channel of information provision is abstract --- the
principal directly controls how much information to release to the agent. In 
\cite{hoffmann2011pre}, the principal produces two independent goods: a
product itself and an additional information provision service. It is
through the consumption of the information provision service that a consumer
learns more about his valuation of the product. Another approach models
information acquisition as a moral hazard problem or an entry problem. For example, in
 \cite{krahmer2011optimal}, the agent can take a
hidden action to gather information; in \cite{lu2021orchestrating}, the agent can incur an entry cost to fully observe the ex post value. Some studies, for example, \cite{armstrong2016search} and \cite{lu2021optimal}, model the information acquisition as searches. In this paper, the buyer's
first-stage consumption plays a {dual} role: The buyer not only enjoys
a payoff but also acquires additional information from the first-stage
consumption. Such information acquisition from the allocation (\textit{i.e.},
consumption) itself differentiates the current work from the above-mentioned
papers.

This paper also features an \emph{intertemporal} problem: The
first-stage allocation not only endogenously shapes the distribution of the
second-stage valuation but also changes the feasibility constraint of the
second-stage allocation. \cite{pavan2014dynamic} accommodate this feature before the current work, but they focus on providing a general approach to tackle dynamic
mechanism design problems. On the other hand, we explicitly characterize the optimum in a
consumer-learning environment.

The rest of the paper is organized as follows. Section \ref{sec:model} sets up the model.
We analyze the solution of a relaxed problem in Section \ref{sec:ProblemOR} and the optimal mechanism in Section \ref{sec:problemo}. Sections \ref{sec:discussion} and \ref{sec:conclude} discuss the results and conclude. The appendix collects some technical proofs.



\section{The Model}\label{sec:model}

A risk-neutral monopolist sells one unit of a \textit{divisible} good to a
risk-neutral buyer in two stages. The buyer's true valuation $V$ of the good
depends on how well the good fits him. At stage one, the buyer is uncertain
about $V$, but he observes a ``rough'' valuation of the good, $v_{1}$.
Relying on this rough valuation, the buyer purchases $q_{1} \in
[0,1]$ units of the good. The true valuation $V$ is jointly determined by $v_1$ and the
additional information $\tilde{v}_{2}$, which is independent of $v_{1}$.\footnote{
The assumption that $v_1$ and $\tilde{v}_{2}$ are independent means that the
ex ante information asymmetry does not depend on the precision of the
second-stage information, which helps us provide a clean characterization of
the optimal mechanism. On the other hand, when considering the more general
setting that $v_1$ and $\tilde{v}_{2}$ are correlated, one has to impose
additional restrictions on the information structure; see, for example, \cite{courty2000sequential} and \cite{esHo2007optimal}. We focus on the current
setting, as working with the more general structure will make the
analysis much more complicated and the new insights less transparent.} We
assume that $E[\tilde{v}_{2}]=0$ and $V=v_{1}+\tilde{v}_{2}$.\footnote{Such an additive form is not an assumption, because one can define the difference between the true valuation and the rough valuation as the additional information.}
Consuming/experiencing $q_{1}$ units of the good provides the buyer with
additional information to learn about $\tilde{v}_{2}$. With a more precise
assessment of the good, at stage two, the buyer decides how many further
units $q_{2}\in \lbrack 0,1-q_{1}]$ to buy. The buyer's outside option is
normalized to be $0$.

From the seller's perspective, $v_{1}$ is a random draw from a cumulative
distribution function (CDF) $G(\cdot )$, which admits a strictly positive
continuous density function $g$ over the support $[0,1]$.\footnote{%
It is without loss to assume that the support is $[0,1]$.} The buyer learns
additional information about $\tilde{v}_{2}$ through consumption: After
buying $q_{1}$ units of the good, the buyer forms a posterior estimate $%
v_{2} $ of $\tilde{v}_{2}$. From an ex ante perspective, $v_{2}$ follows the
CDF $F(\cdot |q_{1})$. The realization of $v_{2}$ is again the buyer's
private information. The seller's goal is to design a contract that
maximizes her revenue.

Intuitively, consuming more at stage one helps the buyer acquire more
precise information at stage two. The precision of the additional
information $v_{2}$ through consumption is captured by the rotation order %
\citep[cf.][]{johnson2006simple, hoffmann2011pre}. Specifically, for any $%
q_{1}\in (0,1]$ and $v_{2}\in (-\infty ,+\infty )$, $F(v_{2}|q_{1})$ is
continuously differentiable in $q_{1}$ such that 
\begin{equation*}
\frac{\partial F(v_{2}|q_{1})}{\partial q_{1}} 
\begin{cases}
> 0, & \text{ when } v_{2} < 0; \\ 
= 0, & \text{ when } v_{2} = 0; \\ 
< 0, & \text{ when } v_{2} > 0.%
\end{cases}%
\end{equation*}
When $q_{1}=0$, $F(\cdot |q_{1})$ degenerates to a mass at $E[\tilde{v}_{2}]
= 0$, capturing no additional information gained if there is no consumption.
For convenience, suppose that when $q_{1}>0$, $F(v_{2}|q_{1})$ is twice
continuously differentiable in $v_2$ and the corresponding density function $%
f(v_{2}|q_{1})>0$. We also assume that for any $v_{2}\neq 0$, $\underset{%
q_{1}\rightarrow 0 + }{\lim }F(v_{2}|q_{1})=F(v_{2}|0)$.



To illustrate the setting, consider the following ``truth-or-noise''
example, which has been widely used in the literature; see, for example, \citet{lewis1994supplying} and \citet{johnson2006simple}.


\begin{exam}[Truth-or-noise]
\label{exam-truth} After consuming $q_1$ units, the buyer observes a signal $%
s$ of $\tilde{v}_{2}$. The signal is true (\textit{i.e.}, $s=\tilde{v}_{2}$)
with probability $q_{1}$, and is completely noisy (\textit{i.e.}, $s$ is an
independent random draw from the same CDF as $\tilde{v}_{2}$) with
probability $1-q_{1}$. Denote the CDF of $\tilde{v}_{2}$ by $H(\cdot )$,
which is twice continuously differentiable over the support $(-\infty,
+\infty )$. Then, when observing the signal $s$, the buyer's posterior
estimate of $\tilde{v}_{2}$ is 
\begin{equation*}
v_{2}=E[\tilde{v}_{2}|s,q_{1}]=q_{1}s+(1-q_{1})E[\tilde{v}_{2}] = q_{1} s.
\end{equation*}
Thus, ex ante, $v_{2}$ follows the CDF $F(v_{2}|q_{1})=H(\frac{v_{2}}{q_{1}}%
) $, which satisfies all the assumptions mentioned above.
\end{exam}

After consuming $q_{1}$ units, imagine that the buyer adopts a simple
threshold plan $v_2$. That is, he will buy the rest $1 - q_{1}$ units at
stage two only when the additional information acquired is sufficiently
good --- \textit{i.e.}, when the additional information is higher than the threshold $v_{2}$. Thus, the buyer's expected consumption in the second stage is $(1-q_{1})(1-F(v_{2}|q_{1}))$,
and the expected total consumption across two stages is 
\begin{equation}
C(v_2,q_1) = q_{1}+(1-q_{1})(1-F(v_{2}|q_{1}))=1-(1-q_{1})F(v_{2}|q_{1}).
\label{def-C-function}
\end{equation}
When $q_{1}>0$, it is clear that a higher threshold $v_{2}$ --- \textit{i.e.}, requiring better information --- leads to a drop in $C$, as the partial derivative with respect to $v_2$,
\begin{equation*}
\frac{\partial C(v_2,q_1)}{\partial v_2} = -(1-q_{1})f(v_{2}|q_{1}) < 0.
\end{equation*}
However, the effect of a higher first-stage consumption $q_{1}$ on $C$ is
ambiguous. The \textquotedblleft marginal rate of
substitution\textquotedblright\ is 
\begin{equation*}
M(v_{2},q_{1}) = \frac{\frac{\partial C(v_2,q_1)}{\partial q_1} }{\frac{\partial C(v_2,q_1)}{\partial v_2}}
= -\frac{1}{1-q_{1}}\frac{F(v_{2}|q_{1})}{f(v_{2}|q_{1})}+\frac{\frac{%
\partial F(v_{2}|q_{1})}{\partial q_{1}}}{f(v_{2}|q_{1})}
\end{equation*}
for any $(v_{2},q_{1}) \in \mathbb{R} \times (0,1)$.

We make the following assumption regarding $M(v_{2},q_{1})$.

\begin{assum}
\label{as:singlecrossing} For any fixed $q_{1}\in (0,1)$ and $%
v_{2}<v_{2}^{\prime }$, 
\begin{equation*}
M(v_{2},q_{1})\leq 0 \quad \Longrightarrow \quad M(v_{2}^{\prime },q_{1})<0.
\end{equation*}
\end{assum}

\begin{rmk}\label{rmk-crossing} Assumption \ref{as:singlecrossing} is a natural assumption
regarding the substitution between the first-stage consumption and the
additional information. It says that: If at a particular level of
first-stage consumption $q_{1}$ and a certain requirement of additional
information $v_{2}$, the buyer is willing to sacrifice his first-stage
consumption in exchange for a lower requirement of information (\textit{i.e.}%
, a lower $v_{2}$), then he will still be willing to do so when the requirement
of information is more stringent than $v_{2}$ (\textit{i.e.}, higher than $%
v_{2}$).

Note that Assumption \ref{as:singlecrossing} holds when $%
F(v_{2}|q_{1})/f(v_{2}|q_{1})$ is increasing in $v_{2}$ and $\frac{\partial
F(v_{2}|q_{1})}{\partial q_{1}}/f(v_{2}|q_{1})$ is decreasing in $v_{2}$.
The former condition is a standard hazard rate assumption. The latter one is
the same as Assumption~3 in \citet{shi2012optimal}, which can be interpreted
as supermodularity. In the truth-or-noise example above, the latter
assumption is automatically satisfied, while the former assumption is
satisfied when $H(x)/h(x)$ is increasing in $x$.
\end{rmk}

Finally, we make the following standard hazard rate assumption about $G$.

\begin{assum}
\label{as:monotonehazardrate} We assume that $\frac{1-G(v_{1})}{g(v_{1})}$
is strictly decreasing in $v_{1}$.
\end{assum}


We shall focus on truthful direct mechanisms 
$\{q_{1}(r_{1}),q_{2}(r_{1},r_{2}),t(r_{1},r_{2})\}_{r_{1}\in [0,1],r_{2}\in \mathbb{R}}$, which is without loss of
generality; see \cite{myerson1986multistage}. In the first stage, when the
buyer reports $r_{1}$, the seller allocates $q_{1}(r_{1})$ units of the good
to him. In the second stage, $v_{2}$ is realized according to $%
F(\cdot |q_{1}(r_{1}))$. Given the buyer's second-stage report $r_{2}$, the
seller allocates $q_{2}(r_{1},r_{2})$ units of the good to the buyer and
demands a payment $t(r_{1},r_{2})$.


\subsection{The buyer's problem}

Suppose that the buyer truthfully reported $v_{1}$ at stage one, but he
reports $r_{2}$ despite that the true second-stage type is $v_{2}$. Let $
\tilde{\pi}(v_{1},r_{2},v_{2})$ be his expected payoff at stage two: 
\begin{equation*}
\tilde{\pi}
(v_{1},r_{2},v_{2})=(v_{1}+v_{2})q_{2}(v_{1},r_{2})-t(v_{1},r_{2}).
\label{e1'}
\end{equation*}
Envelope theorem yields
\begin{equation*}
\frac{\rmd\tilde{\pi}(v_{1},v_{2},v_{2})}{\rmd v_{2}}=\frac{\partial \tilde{\pi}
(v_{1},r_{2},v_{2})}{\partial v_{2}}|_{r_{2}=v_{2}}=q_{2}(v_{1},v_{2}).
\end{equation*}

Denote $\psi (v_{1})=v_{1}-\frac{1-G(v_{1})}{g(v_{1})}$, which is strictly
increasing by Assumption~\ref{as:monotonehazardrate}. The following result
is standard \citep[see, \textit{e.g.},][]{esHo2007optimal}, and its proof is omitted.

\begin{lem}\label{2ndIC} 
\begin{enumerate}
	\item[(i)]
Suppose that the buyer reports the first-stage type $v_{1}$
truthfully. The second-stage IC constraint is satisfied if and only if the following two
conditions hold:
\begin{enumerate}
\item For any $v_{1}$ and $v_{2}$,\footnote{
When $v_{2} < -\psi (v_{1})$, $\int_{-\psi (v_{1})}^{v_{2}}q_{2}(v_{1},s)
\rmd s = - \int_{v_{2}}^{-\psi (v_{1})} q_{2}(v_{1},s)\rmd s$.} 
\begin{equation}
\tilde{\pi}(v_{1},v_{2},v_{2})=\tilde{\pi}(v_{1}, - \psi (v_{1}), - \psi
(v_{1}))+\int_{-\psi (v_{1})}^{v_{2}}q_{2}(v_{1},s)\rmd s.  \label{e2'}
\end{equation}
\item The second-stage allocation $q_{2}(v_{1},v_{2})$ is increasing in $v_{2}$ for any $v_{1}$.
\end{enumerate}

\item[(ii)] On the other hand, suppose that the buyer's first-stage type is $v_{1}$ but he reported $r_{1}$ in the
first stage. Then, when he observes $v_{2}$ in the second stage, he will
report $r_{2}=r_{2}(v_{1},r_{1},v_{2})$ such that 
\begin{equation*}
r_{1}+r_{2}(v_{1},r_{1},v_{2})=v_{1}+v_{2}.
\end{equation*}
\end{enumerate}
\end{lem}


Based on Lemma \ref{2ndIC}, the expected payoff of the buyer with
first-stage type $v_{1}$ and report $r_{1}$ can be expressed as 
\begin{eqnarray*}
U(v_{1},r_{1}) &=&q_{1}(r_{1})\int_{-\infty }^{+\infty }(v_{1}+v_{2})F(\rmd %
v_{2}|q_{1}(r_{1}))  \notag \\
&&+\int_{-\infty }^{+\infty } \left[ 
\begin{array}{c}
(v_{1}+v_{2})q_{2}(r_{1},r_{2}(v_{1},r_{1},v_{2})) \\ 
-t(r_{1},r_{2}(v_{1},r_{1},v_{2}))%
\end{array}%
\right] F(\rmd v_{2}|q_{1}(r_{1})).  \label{1stpayoff}
\end{eqnarray*}
The first-stage IC constraint requires that for any $v_{1}$ and $r_{1}$, 
\begin{equation*}
U(v_{1},v_{1})\geq U(v_{1},r_{1}).
\end{equation*}
The following result provides a necessary condition for the first-stage IC constraint (all the proofs are relegated to the Appendix).


\begin{lem}\label{lem:stage1IC}
\label{1stIC} The first-stage IC constraint implies that for any $v_{1}$ 
\begin{equation*}
U(v_{1},v_{1})=U(0,0)+\int_{0}^{v_{1}}\left[ q_{1}(s)+\int_{-\infty
}^{+\infty }q_{2}(s,v_{2})F(\rmd v_{2}|q_{1}(s))\right] \rmd s.
\end{equation*}
\end{lem}


\subsection{The seller's problem}

The seller's expected revenue is the difference between the social welfare
and the buyer's ex ante expected payoff. By Lemma~\ref{1stIC}, it can be
written as 
\begin{eqnarray*}
R &=&\int_{0}^{1} \left\{ v_{1}q_{1}(v_{1})+\int_{-\infty }^{+\infty
}(v_{1}+v_{2})q_{2}(v_{1},v_{2})F(\rmd v_{2}|q_{1}(v_{1})) \right\} g(v_{1})%
\rmd v_{1} \\
& \quad & -\int_{0}^{1}U(v_{1},v_{1})g(v_{1})\rmd v_{1} \\
&=&\int_{0}^{1}\left[ 
\begin{array}{c}
\psi (v_{1})q_{1}(v_{1}) \\ 
+\int_{-\infty }^{+\infty }[\psi (v_{1})+v_{2}]q_{2}(v_{1},v_{2})F(\rmd %
v_{2}|q_{1}(v_{1}))
\end{array}
\right] g(v_{1})\rmd v_{1}-U(0,0).
\end{eqnarray*}

Now we are ready to state the seller's problem as follows. 
\begin{equation*}
\mbox{Problem (O)}: \qquad \max_{(q_{1},q_{2},t)} R \qquad \qquad
\qquad\qquad\qquad
\end{equation*}%
subject to 
\begin{equation}
\text{constraint (\ref{e2'}) and } q_{2}(v_{1},v_{2}) \text{ is
increasing in } v_{2} \text{ for any } v_{1};  \label{2ndICeq}
\end{equation}%
\begin{equation}
U(v_{1},v_{1})\geq U(v_{1},r_{1}) \text{, for any } v_{1},r_{1};
\label{1stICde}
\end{equation}%
\begin{equation}
0\leq q_{1}(v_{1})\leq 1\text{ and }0\leq q_{2}(v_{1},v_{2})\leq
1-q_{1}(v_{1}) \text{, for any } v_{1},v_{2}.  \label{fesibility}
\end{equation}%
Here, (\ref{2ndICeq}) is the equivalent condition for the second-stage
IC constraint, (\ref{1stICde}) is the first-stage IC constraint, and (\ref{fesibility}) is the
feasibility constraint imposed on allocations.

Clearly, at the optimum, $U(0,0)=0$, and thus the seller's ex ante revenue becomes 
\begin{equation*}
R=\int_{0}^{1}\left[ 
\begin{array}{c}
\psi (v_{1})q_{1}(v_{1}) \\ 
+\int_{-\infty }^{+\infty }[\psi (v_{1})+v_{2}]q_{2}(v_{1},v_{2})F(\rmd 
v_{2}|q_{1}(v_{1}))
\end{array}
\right] g(v_{1})\rmd v_{1}.
\end{equation*} We further drop constraints (\ref
{2ndICeq}) to (\ref{1stICde}) and omit the choice variable $t$ to form a
relaxed problem, Problem~(O-R):
\begin{equation*}
\max_{(q_{1},q_{2})}
R
\end{equation*}
\begin{equation*}
\mbox{ subject to constraint (\ref{fesibility})}.
\end{equation*}

If the solution to Problem~(O-R) also satisfies (\ref{2ndICeq}) and (\ref
{1stICde}), then such a solution also solves Problem~(O). As such, we will
first solve Problem (O-R), and then verify that its solution satisfies all
the constraints in the original problem.



\section{The solution to Problem~(O-R)}\label{sec:ProblemOR}

In Problem (O-R), for each fixed first-stage allocation rule $q_{1}$, the following
second-stage allocation rule $q_2$ obviously maximizes the objective function $R$: 
\begin{equation*}
q_{2}(v_{1},v_{2})=
\begin{cases}
1-q_{1}(v_{1}), & \text{ if } \psi (v_{1})+v_{2}\geq 0; \\ 
0,  & \text{ otherwise.}
\end{cases}
\end{equation*}

Thus, we have\footnote{The detailed derivation can be found in the Appendix.} 
\begin{eqnarray}
R &=&\int_{0}^{1}\left[ \psi (v_{1})q_{1}(v_{1})+\int_{-\psi
(v_{1})}^{+\infty }[\psi (v_{1})+v_{2}](1-q_{1}(v_{1}))F(\rmd %
v_{2}|q_{1}(v_{1}))\right] g(v_{1})\rmd v_{1}  \notag \\
&=&\int_{0}^{1}\left[ \psi (v_{1})+(1-q_{1}(v_{1}))\int_{-\infty }^{-\psi
(v_{1})}F(v_{2}|q_{1}(v_{1}))\rmd v_{2}\right] g(v_{1})\rmd v_{1}.
\label{obj-new}
\end{eqnarray}

To facilitate the presentation, define the seller's revenue from a type-$v_{1} $ buyer with the first-stage consumption $q_1$ as 
\begin{equation}\label{obj-v}
\Pi (q_{1},v_{1})=\psi (v_{1})+(1-q_{1})\int_{-\infty }^{-\psi
(v_{1})}F(v_{2}|q_{1})\rmd v_{2}.  
\end{equation}
For each $v_{1}$, let $q_{1}^{\ast }(v_{1})$ be the maximizer of $\Pi(q_{1},v_{1})$ in $q_{1}\in \lbrack 0,1]$. Finally, define 
\begin{equation*}
v_{1}^{\ast }=\psi ^{-1}(0)\text{ and }\tilde{v}_{1}=\inf \{v_{1}\in \lbrack
0,1]:q_{1}^{\ast }(v_{1})>0\}.  \label{def-cuotoff}
\end{equation*}
We have the following observation.

\begin{lem}\label{lm-zerocutoff} 
The following allocation rule pair $\{q^*_1(v_1),q^*_2(v_1,v_2)\}_{v_1\in [0,1], v_2\in \mathbb{R}}$ solves Problem (O-R): 	
\begin{enumerate}
	\item[(i)] 
	The first-stage allocation $q_{1}^{\ast }(v_1)$ is the maximizer of 
	\begin{equation*}
	\Pi (q_{1},v_{1}) = \psi (v_{1})+(1-q_{1})\int_{-\infty }^{-\psi
		(v_{1})}F(v_{2}|q_{1})\rmd v_{2}
	\end{equation*}
	in $q_{1}\in [0,1]$ for each $v_1$. 
	In particular, there exists a cutoff $\tilde{v}_{1}<v_{1}^{\ast }$ such that  for all $v_{1}\in \lbrack 0,\tilde{v}_{1})$, $q_{1}^{\ast}(v_{1})=0$.
	For all $v_1 \in [\tilde{v}_1,1]$, $q_{1}^{\ast }(v_{1}) \in [0,1)$ and can be characterized by the first-order condition:
		\begin{equation}\label{foc}
		\int_{-\infty }^{-\psi (v_{1})}\left[ -F(v_{2}|q_{1}^{\ast
		}(v_{1}))+(1-q_{1}^{\ast }(v_{1}))\frac{\partial F(v_{2}|q_{1}^{\ast
			}(v_{1}))}{\partial q_{1}}\right] \rmd v_{2}=0.
		\end{equation}
		Moreover, for all $v_1 \in (\tilde{v}_1,1]$, $q^*_1(v_1)\in(0,1)$.
	\item[(ii)] The second-stage allocation rule $q^*_2$ is
	\begin{equation*}
	q^*_{2}(v_{1},v_{2})=
	\begin{cases}
	1-q^*_{1}(v_{1}), & \text{ if } \psi (v_{1})+v_{2}\geq 0; \\ 
	0,  & \text{ otherwise.}
	\end{cases}
	\end{equation*}
		In particular, when  $v_{1}<\tilde{v}_{1}$, $F(\cdot|q^*_1(v_1))$ degenerates to a mass at $0$ and $q^*_2(v_1,0)=0$.
	\end{enumerate}
\end{lem}

The above lemma implies that there is a first-stage cutoff type $\tilde{v}_{1}$, below which both the first-stage allocation and the second-stage allocation are zero. Hence, there is no consumption at the bottom of the distribution $G$. Moreover, this cutoff is strictly below $v_{1}^{\ast }$. This is because a low first-stage value $v_1$ buyer would still be willing to experience the good a bit in the first stage, as he knows that his ex post value (\textit{i.e.}, $v_1+v_2$) is still likely to be high enough. The seller then should take advantage of this by setting a positive allocation for such low first-stage types. However, when the first-stage type is so low (below $\tilde{v}_1$) that the ex post value is quite unlikely to be sufficiently high, the buyer does not find it worthwhile to experience the good; so the seller, anticipating this, should set a zero first-stage allocation in this case.


The following result characterizes the monotonicity of the optimal
first-stage allocation rule.

\begin{lem}\label{lem:q1}
\label{q-monotone} The first-stage allocation $q_{1}^{\ast }(v_{1})$ is
strictly increasing in $v_{1}\in [\tilde{v}_{1},1]$.
\end{lem}

\begin{rmk}
It is worth pointing out that $\Pi (q_{1},v_{1})$ does not have the
supermodularity or single-crossing property in $(q_{1},v_{1})$. As a result,
the standard approach of monotone comparative statics 
\citep[e.g.,][]{milgrom1994monotone}
is not readily
applicable in our setting.
\end{rmk}

\begin{rmk}
\label{ft1} To ease the notations and presentation, we do not take into
account the possibility that for some $v_{1}$, the maximizer $q_{1}^{\ast}(v_{1})$ may not
be unique. This multiplicity issue can be addressed by imposing the following assumption analogous to Assumption \ref{as:singlecrossing}:

\begin{itemize}
\item For any fixed $v_{2}\leq 0$ and any $q_{1},q_{1}^{\prime }\in (0,1)$
with $q_{1}<q_{1}^{\prime }$, 
\begin{equation*}
M(v_{2},q_{1})\leq 0 \qquad \Longrightarrow \qquad M(v_{2},q_{1}^{\prime })
< 0.
\end{equation*}
\end{itemize}

This assumption can be interpreted as a natural substitution condition,
which says that: If at a particular level of first-stage consumption $q_{1}$
and a certain requirement of additional information $v_{2}$, the buyer is
willing to sacrifice his first-stage consumption in exchange for a lower
requirement of information, then he will still be willing to do so when his
first-stage consumption is higher than $q_{1}$. Under this condition, one
can show that any selection of maximizers must be strictly increasing in $%
v_{1}$ when $v_{1}\ge \tilde{v}_{1}$. The proof of this claim is given in the Appendix. 
\end{rmk}


\section{The solution to Problem~(O)}\label{sec:problemo}

\subsection{Optimal direct mechanism}

Having characterized the solution to Problem~(O-R) as in Lemma \ref{lm-zerocutoff}, we can use the envelope conditions in Lemmas \ref{2ndIC} and \ref{1stIC} to construct a payment rule $t^*$, the expression of which is provided in the following result.

\begin{lem}\label{lem:transfer} 
	The payment rule $t^{\ast }$ is specified as follows: 
		\begin{equation*}
			t^{\ast }(v_{1},v_{2}) = 
			\begin{cases}
				(1-q_{1}^{\ast }(v_{1})) p^*_2(v_1) +p_{1}^{\ast }(v_{1}),
				& \mbox{ if } \psi (v_{1}) + v_{2}\geq 0, \\ 
				p_{1}^{\ast }(v_{1}), & \text{ otherwise},
			\end{cases}
		\end{equation*}
		where 
		\begin{align*}
			p_{1}^{\ast }(v_{1}) & =(1-q_{1}^{\ast }(v_{1}))\bigg[ \int_{-\infty
			}^{-\psi(v_{1})}F(v_{2}|q_{1}^{\ast }(v_{1}))\rmd v_{2}-\frac{1-G(v_{1})}{%
				g(v_{1})} \bigg] \\
			& \quad +\int_{0}^{v_{1}}(1-q_{1}^{\ast }(x))F(-\psi (x)|q_{1}^{\ast }(x))%
			\rmd x
		\end{align*}
	   and \begin{equation*}
		p_{2}^{\ast }(v_{1})=\frac{1-G(v_{1})}{g(v_{1})}.
	\end{equation*}
\end{lem}

Thus, we obtain a candidate mechanism $\{q_{1}^{\ast }(v_{1}),q_{2}^{\ast}(v_{1},v_{2}),t^{\ast}(v_{1},v_{2})\}_{v_{1}\in [0,1],v_{2}\in \mathbb{R}}$ for Problem (O). If we can show that  this candidate mechanism satisfies all constraints in Problem (O), then it must solve Problem (O). Clearly, one only needs to verify constraints (\ref{2ndICeq}) and (\ref{1stICde}) --- \textit{i.e.}, the first- and second-stage IC constraints. 

To this end, we begin by considering a menu of \textit{try-and-decide} option contracts $\{p^*_{1}(v_{1}),q^*_{1}(v_{1});p^*_{2}(v_{1})\}_{v_{1}\in [0,1]}$, where functions $p^*_{1}$, $q^*_{1}$, and $p^*_{2}$ are defined in Lemmas \ref{lm-zerocutoff} and \ref{lem:transfer}. The buyer needs to select a contract  from the menu. If for some $r_1 \in [0,1]$, option contract $\{p^*_{1}(r_{1}),q^*_{1}(r_{1});p^*_{2}(r_{1})\}$ is selected, then $p^*_{1}(r_1)$ is the advance payment for a buyer to enter this contract. By paying this advance payment, the buyer not only consumes $q^*_{1}(r_{1})$ units of the good, but also reserves the right to buy the remaining $1 - q^*_{1}(r_{1})$ units at the per-unit strike price $p^*_{2}(r_1)$.

In the Appendix, we show that this menu of contracts implements the above-mentioned direct mechanism. Hence, our direct mechanism satisfies constraints (\ref{2ndICeq}) and (\ref{1stICde}), and thus is a solution to Problem (O), which is the following result.

\begin{prop}\label{prop1} 
	The
	direct mechanism $\{q_{1}^{\ast}(v_{1}),q_{2}^{\ast}(v_{1},v_{2}),t^{\ast }(v_{1},v_{2})\}_{v_1\in [0,1],v_2\in \mathbb{R}}$ can be implemented by a menu of try-and-decide option
	contracts $\{p_{1}^{\ast }(v_{1}),q_{1}^{\ast}(v_{1});p_{2}^{\ast
	}(v_{1})\}_{v_{1}\in [0,1]}$. Hence, this direct mechanism solves Problem (O). 
\end{prop}

What is crucial in the proof of the above proposition is to establish global IC. The argument to establish IC in our problem is non-standard.
In many canonical sequential screening problems in the literature, {\textit{e.g.}}, \cite{courty2000sequential} and \cite{esHo2007optimal},  as long as the allocation rule satisfies certain monotonicity condition in private types, regardless of whether it is the solution of the relaxed problem or not, the allocation rule can be used to construct a direct mechanism satisfying global IC. However, this approach does not work for our problem. Intuitively, the complication arises because the first-stage allocation plays a dual role in our problem --- a higher first-stage consumption is associated with a higher precision of information, but reduces the potential second-stage consumption. 
We sketch the key step of our argument in Section \ref{sec:sketch} and discuss further in Section \ref{sec:discussion}.

\subsection{Sketch of the proof}\label{sec:sketch}

To establish Proposition \ref{prop1}, the two global IC constraints in Problem (O) require that: Facing the menu of contracts
$\{p_{1}^{\ast }(v_{1}),q_{1}^{\ast}(v_{1});p_{2}^{\ast
}(v_{1})\}_{v_{1}\in [0,1]}$, for each $v_1\in [0,1]$, (i) a type-$v_1$ buyer who is under contract $\{p_{1}^{\ast}(v_{1}),q_{1}^{\ast}(v_{1});p_2^{\ast }(v_{1})\}$ will buy the remaining $1-q_{1}^{\ast}(v_{1})$ units if and only if $v_1+v_2\ge p_2^{\ast }(v_{1})$, and (ii) for any $r_1\in [0,1]$, a type-$v_1$ buyer has no strict incentive to choose contract $\{p_{1}^{\ast}(r_{1}),q_{1}^{\ast}(r_{1});p_2^{\ast }(r_{1})\}$ over $\{p_{1}^{\ast}(v_{1}),q_{1}^{\ast}(v_{1});p_2^{\ast }(v_{1})\}$, regardless of his second-stage strategy after choosing $\{p_{1}^{\ast}(r_{1}),q_{1}^{\ast}(r_{1});p_2^{\ast }(r_{1})\}$. Parts (i) and (ii) correspond to the second- and first-stage IC constraints of the direct mechanism, respectively. 
It is easy to see that (i) is trivial. Hence, we focus on (ii). 

Under contract  $\{p_{1}^{\ast}(r_{1}),q_{1}^{\ast}(r_{1});p_2^{\ast }(r_{1})\}$, the optimal second-stage strategy for a type-$v_{1}$ buyer who learns $v_2$ is to buy the remaining portion if and only if $v_2\ge p^*_2(r_1) - v_1$. 
Under the above contract and the optimal second-stage strategy, the type-$v_{1}$ buyer's interim payoff is 
$U(v_{1},r_{1})=w(q^*_{1}(r_{1}),p^*_{2}(r_{1}),v_1)-p^*_1(r_1)$,
where for all $q_1\in [0,1]$, $p_2\ge 0$, and $v_1\in [0,1]$,
\begin{equation*}
w(q_{1},p_{2},v_1)\equiv q_{1}v_{1}+\int_{p_{2}-v_{1}}^{+\infty
}(v_{1}+v_{2}-p_{2})(1-q_{1})F(\rmd v_{2}|q_{1}).
\end{equation*}

It is shown in the Appendix that the difference of interim payoffs between selecting $\{p^*_{1}(v_{1}),q^*_{1}(v_{1}); p^*_{2}(v_{1})\}$ and $\{p^*_{1}(r_{1}),q^*_{1}(r_{1}); p^*_{2}(r_{1})\}$, $\Delta (v_{1},r_{1})=U(v_1,v_1)-U(v_1,r_1)$, is\footnote{In this paper, $w_3$ means partial derivative with respect to the third variable; likewise, $w_{31}$ means the second-order partial derivative with respect to the first and the third variables. Other partial derivatives’ notations are analogous.} 
\begin{align}  \label{eq:doubleintegral}
\int^{v_1}_{r_1} \int^{v_1}_{x} 
\left[ 
\begin{array}{c}
w_{31}(q^*_1(x), p^*_2(x),s){q_1^*}'(x) 
+w_{32}(q^*_1(x), p^*_2(x),s){p_2^*}'(x) 
\end{array}
\right] \rmd s  \rmd x.
\end{align}

Clearly, the first-stage IC constraint holds if and only if $\Delta (v_{1},r_{1}) \geq 0$. To this end, it can be easily shown that $w_{32}(q_{1},p_{2},v_1)\le 0$ for all $q_1\in [0,1]$, $p_2\ge 0$, and $v_1\in [0,1]$. We have established in Lemma \ref{lem:transfer} that $p^*_2(v_1)=\frac{1-G(v_1)}{g(v_1)}$ and thus ${p_2^*}'(\cdot)< 0$ by Assumption \ref{as:monotonehazardrate}. Hence, the second term in the integrand of the double integral (\ref{eq:doubleintegral}), $w_{32}(q^*_1(x), p^*_2(x),s){p_2^*}'(x)\ge 0$. Therefore, it suffices to show that the first term in the integrand is nonnegative.

By Lemmas \ref{lm-zerocutoff} and \ref{lem:q1}, ${q_1^*}'(\cdot)\ge 0$. 
However, the sign of 
\begin{equation*}
w_{31}(q_{1},p_{2},v_1)=F(p_{2}-v_{1}|q_{1})-(1-q_{1})\frac{\partial F(p_{2}-v_{1}|q_{1})}{\partial q_{1}}
\end{equation*}
is ambiguous.
To see this, notice that due to rotation order, when $p_{2}>v_{1}$, $w_{31} > 0$; but when $p_{2}<v_{1}$, $w_{31}$ can be positive or negative. The ambiguity of the sign of $w_{31}$ implies that the sign of $w_{31}(q^*_1(x), p^*_2(x),s){q_1^*}'(x) $ is ambiguous in general when $s$ is between $x$ and $v_1$ in (\ref{eq:doubleintegral}). 
Putting these observations together, the sign of the integrand in double integral (\ref{eq:doubleintegral}) is ambiguous.

We remark that our analysis so far has only used the observation that ${q_1^*}'(\cdot)\ge 0$.  Hence, the ambiguity of the sign of the integrand is not only an issue of our optimal $q_1^*(\cdot)$, but also a problem for more general first-stage allocation rules.

After some transformation, we will obtain that
\begin{equation}\label{eq:intw13}
\int^{v_1}_{r_1} \int^{v_1}_{x} w_{31}(q^*_1(x), p^*_2(x),s){q_1^*}'(x)  \rmd s  \rmd x=
\int^{v_1}_{r_1} {q_1^*}'(x) \int^{x-v_1-\psi(x)}_{-\psi(x)} \xi( y ,q^*_1(x)) \rmd y  \rmd x,
\end{equation}
where $\xi(y,q^*_1(x))\equiv-\frac{\partial C(y,q^*_1(x))}{\partial q^*_1(x)}$.\footnote{Recall that the function $C(\cdot,\cdot)$ is defined in (\ref{def-C-function}).} By the optimality of $q^*_1(\cdot)$ (\textit{i.e.}, the first-order condition (\ref{foc}) which implies that $\int_{-\infty }^{-\psi (x)} \xi(y, q^*_1(x)) \rmd y=0$) as well as Assumption \ref{as:singlecrossing}, we can show that for $v_1\ge r_1$ (resp. $v_1\le r_1$),
$$ \int^{x-v_1-\psi(x)}_{-\psi(x)} \xi( y ,q^*_1(x)) \rmd y \ge 0$$ (resp. $\le 0$) for $x$ between $v_1$ and $r_1$. Hence, the double integral (\ref{eq:intw13}) is nonnegative. 
It is trivial that $$\int^{v_1}_{r_1} \int^{v_1}_{x} w_{32}(q^*_1(x), p^*_2(x),s){p_2^*}'(x) \rmd s  \rmd x\ge 0.$$   
Combining the two inequalities above, one can conclude that $\Delta (v_{1},r_{1})\ge 0$. 


\subsection{Implementation}

The following lemma summarizes several useful
properties of the payment rule, and we will discuss them after Proposition \ref{propimplementation}.

\begin{lem}\label{lem:payment}
	\begin{enumerate}
		\item[(i)] For $v_1<\tilde{v}_1$, $p_{1}^{\ast }(v_{1})=0$ and $t^*(v_{1},0)=0$. 
		
		\item[(ii)] The first-stage payment $p_{1}^{\ast }(v_{1})$ is strictly increasing in $v_{1}\in [ \tilde{v}_{1},1]$ and equal to  $q^*_1(\tilde{v}_1)\frac{1-G(\tilde{v}_1)}{g(\tilde{v}_1)}\ge 0$ for $v_{1} =\tilde{v}_{1}$.
		
		\item[(iii)] The second-stage payment $p_{2}^{\ast }(v_{1})$ is strictly decreasing
		in $v_1\in [0, 1]$ and equal to zero when $v_1=1$.
		
		\item[(iv)] The function $p_{1}^{\ast }(v_{1}) + (1-q_{1}^{\ast }(v_{1}))
		p_{2}^{\ast }(v_{1})$ is strictly decreasing in $v_{1}\in  [0,1]$.
		
		\item[(v)] The function $p_{1}^{\ast }(v_{1}) + (1-q_{1}^{\ast }(v_{1}))p_{2}^{\ast }(v_{1})(1-F(-\psi(v_1)|q^*_1(v_1)))
		$ is strictly increasing in $v_{1}\in \lbrack  \tilde{v}_{1},1]$ and equal to zero elsewhere.
	\end{enumerate}
\end{lem}

\begin{rmk}[No participation at the bottom]
	According to Lemmas \ref{lm-zerocutoff} and \ref{lem:payment} (i), when  $v_{1}<\tilde{v}_{1}$, there is zero consumption in both stages and the total payment is zero. As such, the buyer with the first-stage type lower than $\tilde{v}_{1}$ is completely shut down --- he does not participate.\label{rmk:noparticipation} 
\end{rmk}

According to Remark \ref{rmk:noparticipation}, it is immediate that the ``reduced" menu of try-and-decide option contracts $\{p_{1}^{\ast }(v_{1}),q_{1}^{\ast}(v_{1});p_{2}^{\ast}(v_{1})\}_{v_{1}\in [\tilde{v}_1,1]}$ implements the solution of Problem (O), since a buyer with first-stage type $v_{1}<\tilde{v}_{1}$ simply does not participate. Hence, we have the following result.

\begin{prop}\label{propimplementation} 
	The solution of Problem (O) can be implemented by a menu of try-and-decide option
	contracts $\{p_{1}^{\ast }(v_{1}),q_{1}^{\ast}(v_{1});p_{2}^{\ast
	}(v_{1})\}_{v_{1}\in [\tilde{v}_1,1]}$. 
\end{prop}

According to Lemmas \ref{q-monotone} and \ref{lem:payment},
the first-stage payment $p_{1}^{\ast}(v_{1})$ and consumption $q_{1}^{\ast }(v_{1})$ are strictly increasing when $v_{1}\geq \tilde{v}_1$. However, the per-unit strike price $p_{2}^{\ast}(v_1)$ is strictly decreasing, and the total payment conditional on buying the entire portion of the good --- \textit{i.e.}, $p_{1}^{\ast}(v_{1})+(1-q^*_1(v_1))p_{2}^{\ast}(v_1)$ --- is strictly decreasing
in $v_{1}$. 
This implies that in equilibrium, a buyer with a higher first-stage type will choose a contract with a higher advance payment and higher first-stage consumption, in exchange for a lower per-unit strike price for additional consumption in the second stage and a lower cost for purchasing the entire unit of the good.

The intuition is clear.
For a high $v_{1}$-type buyer, he is more confident that his ex post
valuation of the good, $v_{1}+v_{2}$, is sufficiently high so that he will likely
end up buying the entire good. The buyer is thus incentivized to choose a contract that ``secures" a large first-stage consumption and first-stage payment so that he can enjoy a lower per-unit strike price in stage two and a lower cost for purchasing the entire unit.
However, this will be quite risky for a low $v_{1}$-type. If he does
so, in spite of a lower per-unit second-stage price, he has to pay to consume more in
the first stage. Yet, since his
first-stage type is low, he really wants to experience the good a bit to
make a better decision in the second stage, rather than \textquotedblleft
blindly\textquotedblright\ having a high first-stage consumption level,
which can lead to a rather low expected payoff given his low first-stage
type.

This intuition echoes phenomena seen in practice. For example, consumers who are optimistic about their matching quality with a car tend to buy the car outright, because buying the car outright is usually cheaper than leasing it with the intention to buy out the car at lease end. However, those who are not as optimistic may choose to lease the car first before making a buyout decision. This practice can be expensive, but gives consumers an opportunity to experience the car before committing to it for a longer duration.

Lemma \ref{lem:payment}~(v) implies that $\Pi(q^*_1(v_1),v_1)$ is strictly increasing in $v_1$ for $v_1\in [\tilde{v}_1,1]$. Namely, the seller can expect to extract more surplus from a higher first-stage type.


\subsection{An illustrative example}\label{sec:illustrativeexample}

\label{subsec-illustrative}

In this section, we provide an illustrative example, which demonstrates the
optimal mechanism that is identified in Proposition~\ref{prop1}. In
addition, we shall also illustrate that the monotonicity of the first-stage allocation rule does
not imply global IC. To be precise, we explicitly construct an increasing
first-stage allocation rule $\hat{q}_{1}$, and show that it cannot be part of an
incentive-compatible mechanism.

Recall the truth-or-noise model in Example~\ref{exam-truth}. Let $G$ be the
uniform distribution on $[0, 1]$ with the constant density $g \equiv 1$, and 
$H$ be the normal distribution $N(0, 1)$ with mean $0$ and variance $1$ (%
\textit{i.e.}, the density is $h(v_2) = \frac{1}{\sqrt{2 \pi}} e^{- \frac{%
v_2^2}{2}}$). Then $\psi (v_{1}) = 2 v_{1} - 1$, 
\begin{equation*}
F(v_{2}|q_{1}) = H(\frac{v_{2}}{q_{1}}) = \int_{-\infty}^{\frac{v_{2}}{q_{1}}%
} \frac{1}{\sqrt{2 \pi}} e^{- \frac{s^2}{2}} \rmd s, \quad \mbox{ and }
\quad f(v_{2}|q_{1}) = \frac{1}{q_{1} \sqrt{2 \pi}} e^{- \frac{v_{2}^2}{2
q^2_{1}}}.
\end{equation*}
In the Appendix, we verify that $F$ satisfies the rotation order, and
Assumptions~\ref{as:singlecrossing} and \ref{as:monotonehazardrate} hold.

\begin{figure}[]
\centering
\begin{minipage}[b]{0.35\textwidth}
		\includegraphics[width = \textwidth]{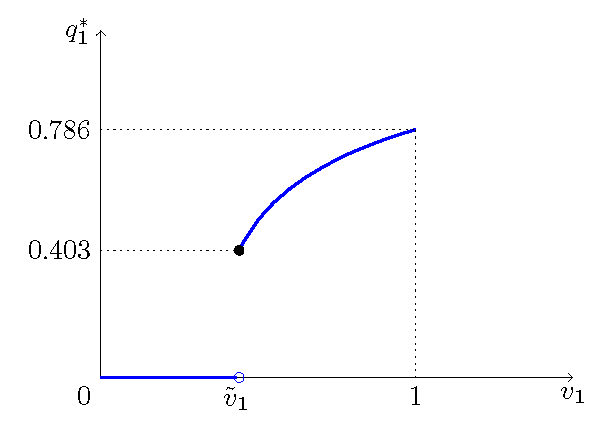}
		\caption{$q_1^*$}
		\label{fig-1}
	\end{minipage}
\hfill 
\begin{minipage}[b]{0.62\textwidth}
		\includegraphics[width = \textwidth]{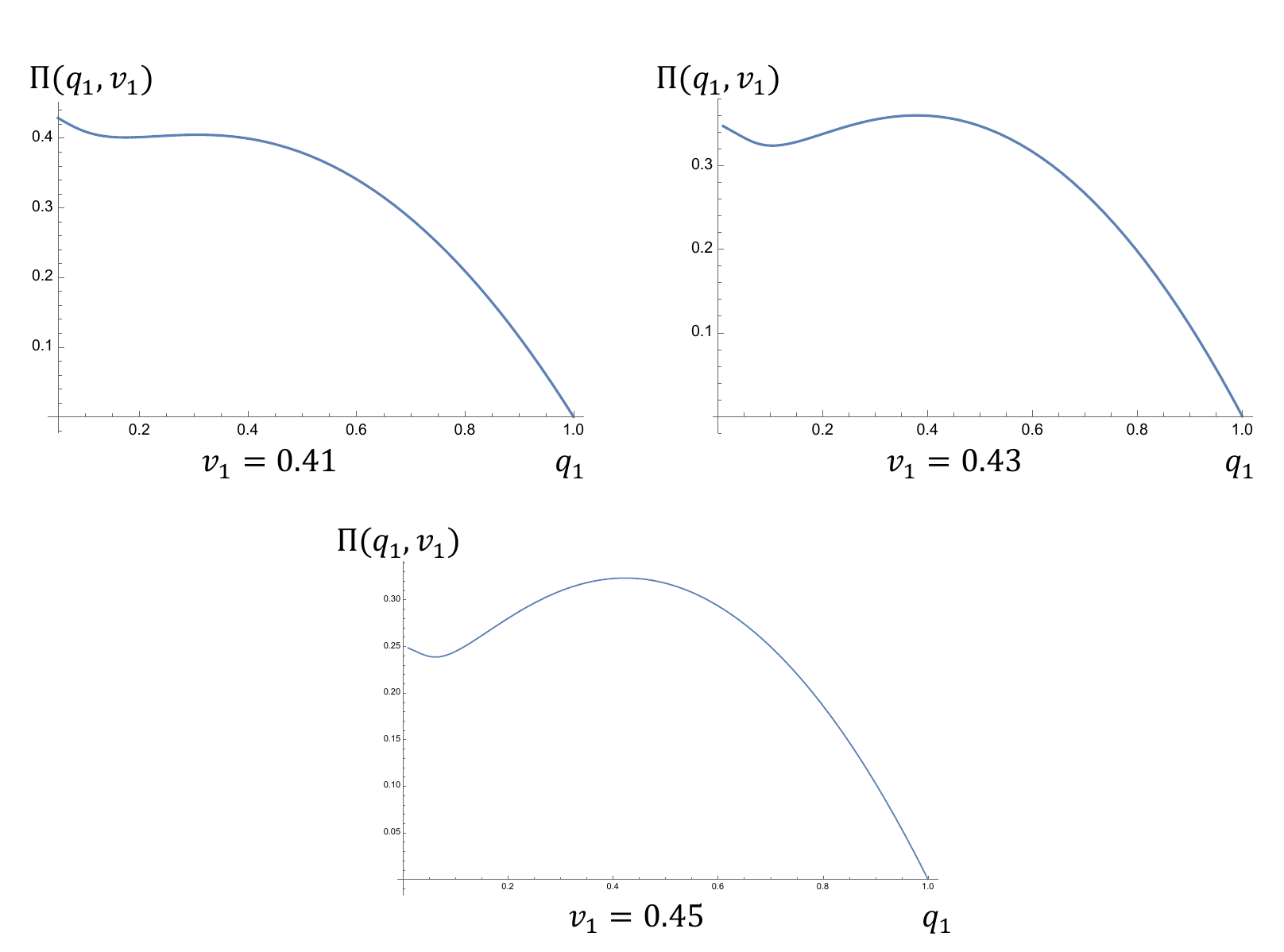}
		\caption{$\Pi(q_1, v_1)$}
		\label{fig-2}
	\end{minipage}
\end{figure}

As shown in Proposition~\ref{prop1}, the optimal first-stage allocation rule $
q_1^*$ must maximize 
\begin{equation*}
(1-q_{1})\int_{-\infty }^{-\psi (v_{1})}F(v_{2}|q_{1})\rmd v_{2}.
\end{equation*}
Figure~\ref{fig-1} numerically illustrates the optimal allocation rule.%
\footnote{In this example, we plot the figures by simulation.}
In this example, $%
\tilde{v}_1 \approx 0.43 < 0.5 = v_1^*$. In Figure~\ref{fig-2}, we plot $%
\Pi(q_1, v_1)$ when $v_1 = 0.41$, $0.43$, and $0.45$. As can be seen, $%
q_1^*(v_1)$ is unique and higher than $q_1^*(\tilde{v}_1) \approx 0.403$
when $v_1 > \tilde{v}_1 \approx 0.43$, is either $0$ or almost $0.403$ when $%
v_1 = \tilde{v}_1$, and is $0$ when $v_1 < \tilde{v}_1$. This pattern
explains the jump of $q_1^*$ at $\tilde{v}_1$ in Figure~\ref{fig-1}. 



Below, we construct another allocation rule $(\hat{q}_1$, $\hat{q}_2)$: 
\begin{equation*}
\hat{q}_1 (v_1) = 
\begin{cases}
\psi^2 (v_1), & v_1 \ge \frac{1}{2}, \\ 
0, & \mbox{otherwise};%
\end{cases}
\qquad \hat{q}_2 (v_1, v_2) = 
\begin{cases}
1 - \hat{q}_1 (v_1), & \psi (v_{1})+v_{2} \geq 0, \\ 
0, & \mbox{ otherwise}.%
\end{cases}%
\end{equation*}
It is clear that $\hat{q}_1$ is increasing. In the Appendix, we show that $(\hat{q}_1$, $\hat{q}_2)$ cannot be the allocation rule in an
incentive-compatible mechanism.


\section{Discussions}\label{sec:discussion}

\subsection{Monotonicity does not imply global IC}
In many canonical sequential screening problems in the literature \citep[cf.][]{courty2000sequential, esHo2007optimal},  as long as the allocation rule satisfies certain monotonicity condition, regardless of whether it is the solution of the relaxed problem, the allocation rule can be used to construct a mechanism satisfying global IC. As such, the standard treatment in the literature focuses on identifying sufficient conditions under which the solution of the relaxed problem is monotone. However, this approach does not work in the current paper and we will discuss our approach by connecting it with the literature.

By nicely linking a canonical sequential screening problem to a static screening problem, \cite{krahmer2017sequential} show that the first-order stochastic dominance (FOSD) ranking of first-stage types in a canonical sequential screening problem as in \cite{courty2000sequential} corresponds to the single-crossing condition\footnote{The single-crossing condition is also refereed to as the Spence-Mirrlees condition or the constant sign condition in the literature. Note that this condition should not be confused with the one required by our Assumption \ref{as:singlecrossing}. To minimize confusion, we use the term single-crossing condition to refer to the standard condition imposed by canonical screening problems exclusively.} in the corresponding static screening problem; and conversely, a sequential screening problem without FOSD corresponds to a static screening problem without the single-crossing condition.
With FOSD, the above-mentioned standard treatment in sequential screening problems --- \textit{i.e.}, finding sufficient conditions under which the optimal allocation rule in a certain relaxed problem is monotone --- works for establishing global IC. This is similar to the well-known result that in static screening problems, the single-crossing condition ensures that local IC plus monotonicity of the allocation rule implies global IC. 

However, the counterpart of the single-crossing condition does not hold in our corresponding static screening problem, as types are not ranked by FOSD here. More specifically, in the corresponding static screening problem, the condition requires that $w_{31}$ has a constant sign in the respective integration region in our optimal solution, but as we have discussed in Section \ref{sec:sketch}, $w_{31}$ does not satisfy this property.\footnote{The lack of single-crossing condition issue has been discussed by \cite{araujo2010adverse} and \cite{schottmuller2015adverse} in their respective static screening environments.} Due to the lack of the single-crossing condition, rather than just establishing the monotonicity of the allocation rule as in canonical sequential screening problems, we need to deal with global IC in a non-standard way; indeed, the monotonicity of allocation rule does not imply global IC, as seen in the example in Section \ref{sec:illustrativeexample}. In particular, as is illustrated in Sections \ref{sec:sketch} and \ref{sec:illustrativeexample}, the optimality of the first-stage allocation rule, \textit{i.e.}, the fact that the first-stage allocation rule solves the relaxed problem, is explicitly used in establishing global IC, and this contrasts with many canonical sequential screening problems.

We remark that the expression of $w_{31}$ is complicated. In particular, the cutoff point at which $w_{31}$ changes sign depends on $v_1$, $p_2$, and $q_1$ simultaneously.
This is due to the \textit{dual} role played by the first-stage allocation. First, the first-stage allocation is a device of \textit{information acquisition}. Second, it defines an \textit{intertemporal} problem: It affects the feasibility constraint of the second-stage allocation, as the second-stage allocation cannot exceed the remaining portion of the good. 
In problems where the first-stage allocation only affects information acquisition \citep[cf.][]{hoffmann2011pre}, the counterpart of $w_{31}$ has a more clear-cut structure --- the point at which their $w_{31}$ changes sign depends on $v_1$ and $p_2$ only. The dual role of our first-stage allocation leads to new observations as we will discuss in Sections \ref{sec:monotonicity} and \ref{sec:distortion}.


\subsection{Monotonicity of the first-stage allocation rule}\label{sec:monotonicity}
Notice that at the optimum, a higher first-stage type always consumes more,
and thus he acquires more information. This is a bit
counter-intuitive. After all, a sufficiently high type does not need to
actively acquire additional information, because his current information
(first-stage type) is already sufficiently good; similarly, a
sufficiently low type also does not want to incur a cost to acquire
information, because his current information is already so bad that costly
information acquisition is not beneficial. Thus, intuitively, only
\textquotedblleft middle\textquotedblright\ types have a strong incentive to
acquire information. In fact, this is indeed the case in the continuous information acquisition model of  \cite{hoffmann2011pre} and the discrete information acquisition model of \cite{krahmer2011optimal}. In particular, 
\citet{hoffmann2011pre} study a similar problem as our paper. They show that the level of
information acquisition is of hump shape: The level first increases and then decreases in the first-stage type.

The crucial difference is that in our paper, the first-stage
consumption plays a \textit{dual} role for the buyer. On the one hand, it is
a device for acquiring more information that shapes the distribution of
additional information as in these papers; on the other hand, it also 
\textit{directly} changes the buyer's payoff by determining the division
between the first- and second-stage consumption. The latter role is absent
in these papers.

\subsection{Distortion}\label{sec:distortion}
Finally, we compare the consumption levels in both stages with their counterparts in the first-best
scenario, in which the buyer's first- and second-stage types are public.
Denote the buyer's first- and second-stage types as $v_{1}$ and $v_{2}$,
respectively. Let $q_{1}^{FB}(v_{1})$ and $q_{2}^{FB}(v_{1},v_{2})$ be his socially efficient
first- and second-stage consumption, respectively. The expected social surplus from a
buyer with first-stage type $v_1$ is
\begin{equation*}
v_{1}q_{1}^{FB}(v_{1})+\int_{-\infty }^{+\infty
}(v_{1}+v_{2})q_{2}^{FB}(v_{1},v_{2})F(\rmd v_{2}|q_{1}^{FB}(v_{1})).
\end{equation*}
Suppose $
q_{1}^{FB}$ is given. Then, the first-best $
q_{2}^{FB}$ is
\begin{equation*}
q_{2}^{FB}(v_{1},v_{2})=
\begin{cases}
1-q^{FB}_{1}(v_{1}), & \text{ if } v_{1}+v_{2}\geq 0; \\ 
0,  & \text{ otherwise.}
\end{cases}
\end{equation*}

Thus, the above expression of social surplus can be written as\footnote{The derivation is almost the same as that for the integral in (\ref{obj-new}); one just needs to replace $\psi (v_{1})$ and $q_{1}(v_{1})$ there with $
v_{1}$ and $q_{1}^{FB}(v_{1})$, respectively.}
\begin{eqnarray*}
&&v_{1}q_{1}^{FB}(v_{1})+\int_{-v_{1}}^{+\infty
}(v_{1}+v_{2})(1-q_{1}^{FB}(v_{1},v_{2}))F(\rmd v_{2}|q_{1}^{FB}(v_{1})) \\
&=&v_{1}+(1-q_{1}^{FB}(v_{1}))\int_{-\infty
}^{-v_{1}}F(v_{2}|q_{1}^{FB}(v_{1}))\rmd v_{2}.
\end{eqnarray*}%
Therefore, $q_{1}^{FB}(v_{1})$ maximizes the function 
\begin{equation*}
\tilde{\Pi}(q_{1},v_{1})=v_{1}+(1-q_{1})\int_{-\infty
}^{-v_{1}}F(v_{2}|q_{1})\rmd v_{2}.
\end{equation*}%
Comparing with (\ref{obj-v}), one can see that $q_{1}^{FB}(\psi(v_{1}))=q_{1}^{
\ast }(v_{1})$ for all $v_1\in [v^*_1,1]$. 
This leads to $q_{1}^{FB}(v_{1})=q_{1}^{\ast }(\psi^{-1} (v_{1}))$ for all $v_1\in [0,1]$. By Lemma \ref{q-monotone}, the fact that $\psi$ is strictly increasing, and the fact that $\psi^{-1}(0)=v^*_1>\tilde{v}_1$, $q_{1}^{FB}$ is strictly increasing. Also, recall that $v_1\ge \psi(v_1)$ with strict inequality when $v_{1}\in [0,1)$,
we have $q_{1}^{FB}(v_{1})\ge q_{1}^{FB }(\psi(v_{1}))=q_{1}^{\ast }(v_{1})$ with strict inequality when $v_{1}\in [0,1)$. We summarize these observations below.

\begin{prop}
\label{prop-fb}
\begin{enumerate}
	\item[(i)] $q_{1}^{FB}(v_{1})=q_{1}^{\ast }(\psi^{-1} (v_{1}))$ for all $
	v_1\in [0,1]$;
	\item[(ii)] $q_{1}^{FB}(v_{1})$ is strictly increasing when $v_{1}\in [0,1]$; 
	\item[(iii)] $q_{1}^{FB}(v_{1})\geq q_{1}^{\ast }(v_{1})$, with strict inequality when $v_{1}\in [0,1)$.
\end{enumerate}
\end{prop}

The above proposition implies that in general, there is an under-provision of
information when asymmetric information arises compared to the
first best. This again differs from \citet{hoffmann2011pre}
and  \citet{krahmer2011optimal}, in which there can be an over- and under-provision of information, depending on the first-stage type.

Now let us focus on the case that $v_{1}\in [0,1)$ to compare the optimal and the first-best second-stage allocations. 
For $v_2 \ge -\psi(v_1)$, $q^{FB}_2(v_1,v_2)< q^{*}_2(v_1,v_2)$ and $q^{FB}_1(v_1)+q^{FB}_2(v_1,v_2)= q^{*}_1(v_1)+q^{*}_2(v_1,v_2)=1$; 
for $v_2\in [-v_1,-\psi(v_1))$, $q^{FB}_2(v_1,v_2)>q^{*}_2(v_1,v_2)=0$ and $q^{FB}_1(v_1)+q^{FB}_2(v_1,v_2)=1> q^{*}_1(v_1)+q^{*}_2(v_1,v_2)$;
for $v_2 < -v_1$, $q^{FB}_2(v_1,v_2)= q^{*}_2(v_1,v_2)=0$ and $q^{FB}_1(v_1)+q^{FB}_2(v_1,v_2)> q^{*}_1(v_1)+q^{*}_2(v_1,v_2)$. This means that although the second-stage consumption can be upward or downward distorted, the total consumption can only be downward distorted.

\section{Concluding Remarks}\label{sec:conclude}

In this paper, we study the two-stage revenue-maximizing mechanism when the
buyer acquires additional information by first-stage consumption. The
buyer's decision of first-stage consumption depends on his private, prior
valuation of the good. A higher first-stage consumption level leads to a
more precise value estimate of the good but reduces the
available amount of consumption left for the second stage. The key feature of our model is that the
first-stage consumption plays a dual rule: The buyer not only enjoys a
payoff but also acquires additional information from the first-stage
consumption.  

We fully characterize the optimum and find that it can be implemented by a menu of try-and-decide option contracts,
consisting of a first-stage price-quantity pair and a second-stage per-unit
price for the remaining quantity. A larger first-stage quantity is paired
with a higher first-stage price but a lower second-stage per-unit price. In
equilibrium, a higher first-stage valuation buyer pays more to have higher
first-stage consumption and enjoys a lower second-stage price.

Since the second-stage type's distribution is not ranked by first-order
stochastic dominance, we face the difficulty of the failure of the
single-crossing condition when establishing global IC. The monotonicity in the first-stage consumption plus local IC is 
not sufficient for global IC.
As such, we cannot apply the usual approach as in many dynamic mechanism
design papers, which assume FOSD, to establish global IC.

In our current analysis, we assumed that the second-stage type's
distribution only depends on the first-stage consumption. A more general
environment is when it depends on both the first-stage consumption level and the
first-stage type. This is a highly meaningful but challenging direction to
explore. We leave it for future work.

{\singlespacing
\addcontentsline{toc}{section}{References} 
\bibliographystyle{chicago}
\bibliography{bib}
}


\newpage

\appendix

\section{Appendix}



\medskip

\subsection{Proof of Lemma \ref{1stIC}}

By Lemma \ref{2ndIC}, the expected payoff of the buyer with first-stage type 
$v_{1}$ and report $r_{1}$ can be expressed as 
\begin{align*}
U(v_{1},r_{1})=& q_{1}(r_{1})\int_{-\infty }^{+\infty }(v_{1}+v_{2})F(\rmd %
v_{2}|q_{1}(r_{1}))  \notag \\
& +\int_{-\infty }^{+\infty }\left[ 
\begin{array}{c}
(v_{1}+v_{2})q_{2}(r_{1},r_{2}(v_{1},r_{1},v_{2})) \nonumber \\ 
-t(r_{1},r_{2}(v_{1},r_{1},v_{2}))%
\end{array}%
\right] F(\rmd v_{2}|q_{1}(r_{1})).  \notag \\
=&q_{1}(r_{1})v_{1}+\int_{-\infty }^{+\infty }\left[ 
\begin{array}{c}
(v_{1}+v_{2})q_{2}(r_{1},r_{2}(v_{1},r_{1},v_{2})) \\ 
-t(r_{1},r_{2}(v_{1},r_{1},v_{2}))%
\end{array}%
\right] F(\rmd v_{2}|q_{1}(r_{1})),  
\end{align*}
where the second equality uses the fact that $E[v_{2}]=0$. Taking the
partial derivative with respect to $v_{1}$ leads to\footnote{
The (almost everywhere) differentiability of $q_{2}$ and $t$ in $r_{2}$
follows from the second-stage IC constraint.}
\begin{equation*}
\frac{\partial U(v_{1},r_{1})}{\partial v_{1}}=q_{1}(r_{1})+\int_{-\infty
}^{+\infty }\left\{ 
\begin{array}{c}
q_{2}(r_{1},r_{2}(v_{1},r_{1},v_{2})) \\ 
+\frac{\partial r_{2}(v_{1},r_{1},v_{2})}{\partial v_{1}}\left[ 
\begin{array}{c}
(v_{1}+v_{2})\frac{\partial q_{2}(r_{1},r_{2}(v_{1},r_{1},v_{2}))}{\partial
r_{2}} \\ 
-\frac{\partial t(r_{1},r_{2}(v_{1},r_{1},v_{2}))}{\partial r_{2}}%
\end{array}%
\right]%
\end{array}%
\right\} F(\rmd v_{2}|q_{1}(r_{1})).
\end{equation*}

Since$\ r_{2}(v_{1},r_{1},v_{2})$ is the optimal report following a lie in
the first stage --- \textit{i.e.}, it maximizes the second-stage expected
payoff after a lie --- it must satisfy the first-order condition, so 
\begin{equation*}
(v_{1}+v_{2})\frac{\partial q_{2}(r_{1},r_{2}(v_{1},r_{1},v_{2}))}{\partial
r_{2}}-\frac{\partial t(r_{1},r_{2}(v_{1},r_{1},v_{2}))}{\partial r_{2}}=0.
\end{equation*}%
Therefore,%
\begin{equation*}
\frac{\partial U(v_{1},r_{1})}{\partial v_{1}}=q_{1}(r_{1})+\int_{-\infty
}^{+\infty }q_{2}(r_{1},r_{2}(v_{1},r_{1},v_{2}))F(\rmd v_{2}|q_{1}(r_{1})).
\end{equation*}%
Note that $r_{2}(v_{1},r_{1},v_{2})=v_{2}$ when $r_{1}=v_{1}$ (\textit{i.e.}%
, a truthful report in the first stage). The envelope theorem %
\citep[cf.][]{milgrom2002envelope} implies that 
\begin{equation*}
U(v_{1},v_{1})=U(0,0)+\int_{0}^{v_{1}}\left[ q_{1}(s)+\int_{-\infty
}^{+\infty }q_{2}(s,v_{2})F(\rmd v_{2}|q_{1}(s))\right] \rmd s.
\end{equation*}%
This completes the proof. \ $\square$

\medskip


\subsection{Derivation of Equation (\ref{obj-new})}

Before proving Equation~(\ref{obj-new}), we first prove the following preparatory lemma.

\begin{lem}
	\label{lem-parts} For any $a \in (-\infty, + \infty)$ and $q_1 > 0$, 
	\begin{equation*}
	\int_{-\infty}^{a} v_2 F(\rmd v_2 | q_1) = a F(a | q_1) - \int_{-\infty}^{a}
	F(v_2 | q_1) \rmd v_2.
	\end{equation*}
\end{lem}

\begin{proof}
	We first claim that as $b \to -\infty$, $b F(b | q_1) \to 0$. Suppose that the claim is not true. Then there exists some $\epsilon > 0$ and a negative sequence $\{b_k\}$ that converges to $-\infty$ such that $|b_k| F(b_k | q_1) > \epsilon$ for any $k$. Since the integral $\int_{-\infty}^{+\infty} v_2 F(\rmd v_2| q_1)$ is well defined, there exists some sufficiently large $K$ such that for any $k \ge K$, $\int_{-\infty}^{b_k} |v_2| F(\rmd v_2| q_1) < \epsilon$. It implies that 
	$$ |b_k| F(b_k| q_1) = \int_{-\infty}^{b_k} |b_k| F(\rmd v_2| q_1) \le \int_{-\infty}^{b_k} |v_2| F(\rmd v_2| q_1) < \epsilon,
	$$
	which is a contradiction.

	For any $b < \min\{a, 0\}$, due to integral by parts,
	$$\int_{b}^{a} v_2 F(\rmd v_2 | q_1) = a F(a | q_1) - b F(b | q_1) - \int_{b}^{a} F(v_2 | q_1) \rmd v_2.
	$$
	It implies that 
	$$\int_{b}^{a} F(v_2 | q_1) \rmd v_2 = a F(a | q_1) - b F(b | q_1) - \int_{b}^{a} v_2 F(\rmd v_2 | q_1).
	$$
	It is obvious that $\int_{b}^{a} F(v_2 | q_1) \rmd v_2$ is decreasing in $b$. In addition, it is bounded as $b F(b | q_1) \to 0$ and $\int_{b}^{a} v_2 F(\rmd v_2 | q_1) \to \int_{-\infty}^{a} v_2 F(\rmd v_2 | q_1)$ when $b \to -\infty$. Thus, the limit $\lim_{b \to -\infty} \int_{b}^{a} F(v_2 | q_1) \rmd v_2$ exists, which is $\int_{-\infty}^{a} F(v_2 | q_1) \rmd v_2$. This completes the proof.
\end{proof}

Now we derive Equation~(\ref{obj-new}).
\begin{align*}
& q_{1}(v_{1})\psi (v_{1})+(1-q_{1}(v_{1}))\int_{-\psi (v_{1})}^{+\infty
}[\psi (v_{1})+v_{2}]F(\rmd v_{2}|q_{1}(v_{1})) \\
=& q_{1}(v_{1})\psi (v_{1})+(1-q_{1}(v_{1}))\left[ \psi (v_{1})[1-F(-\psi
(v_{1})|q_{1}(v_{1}))]+\int_{-\psi (v_{1})}^{+\infty }v_{2}F(\rmd %
v_{2}|q_{1}(v_{1}))\right] \\
=& \psi (v_{1})-(1-q_{1}(v_{1}))\left[ \psi (v_{1})F(-\psi
(v_{1})|q_{1}(v_{1}))+\int_{-\infty }^{-\psi (v_{1})}v_{2}F(\rmd %
v_{2}|q_{1}(v_{1}))\right] \\
=& \psi (v_{1})-(1-q_{1}(v_{1}))\left[ 
\begin{array}{c}
\psi (v_{1})F(-\psi (v_{1})|q_{1}(v_{1}))  \\  +
\left( -\psi (v_{1})F(-\psi (v_{1})|q_{1}(v_{1}))-\int_{-\infty }^{-\psi
(v_{1})}F(v_{2}|q_{1}(v_{1}))\rmd v_{2}\right)%
\end{array}%
\right] \\
=& \psi (v_{1})+(1-q_{1}(v_{1}))\int_{-\infty }^{-\psi
(v_{1})}F(v_{2}|q_{1}(v_{1}))\rmd v_{2},
\end{align*}%
where the second equality uses the fact that 
\begin{equation*}
\int_{-\psi (v_{1})}^{+\infty }v_{2}F(\rmd v_{2}|q_{1}(v_{1})) +
\int_{-\infty }^{-\psi (v_{1})}v_{2}F(\rmd v_{2}|q_{1}(v_{1})) = 0,
\end{equation*}
and the third equality holds because of Lemma~\ref{lem-parts}. $\qquad
\qquad \square$ 

\medskip


\subsection{Proof of Lemma~\ref{lm-zerocutoff}}


We first establish (i). Since the optimal $q^*_1(\cdot)$ maximizes (\ref{obj-v}), the solution either satisfies the first-order condition or is the corner solution.

When $q_{1}>0$, 
\begin{equation*}
\frac{\partial \Pi (q_{1},v_{1})}{\partial q_{1}}=-\int_{-\infty }^{-\psi
(v_{1})}F(v_{2}|q_{1})\rmd v_{2}+(1-q_{1})\int_{-\infty }^{-\psi (v_{1})}%
\frac{\partial F(v_{2}|q_{1})}{\partial q_{1}}\rmd v_{2}.
\end{equation*}%

We first show that $q_{1}=1$ cannot be optimal. In fact, 
\begin{equation*}
\frac{\partial \Pi (q_{1},v_{1})}{\partial q_{1}}|_{q_{1}=1}=-\int_{-\infty
}^{-\psi (v_{1})}F(v_{2}|1)\rmd v_{2}<0,
\end{equation*}%
which makes $q_{1}=1$ suboptimal.

On the other hand, if $q_{1}=0$ is optimal, it must be the case that $\psi
(v_{1})<0$ --- \textit{i.e.}, $v_{1}<v_{1}^{\ast }$. In fact, when $\psi
(v_{1})\geq 0$ and $q_{1}=0$, the seller's revenue is 
\begin{equation*}
\Pi (0,v_{1})=\psi (v_{1})+\int_{-\infty }^{-\psi (v_{1})}F(v_{2}|0)\rmd %
v_{2}=\psi (v_{1})\text{,}
\end{equation*}%
which is strictly dominated by, for example, choosing $q_{1}=\frac{1}{2}$:
\begin{equation*}
\Pi (\frac{1}{2},v_{1})=\psi (v_{1})+\frac{1}{2}\int_{-\infty }^{-\psi
(v_{1})}F(v_{2}|\frac{1}{2})\rmd v_{2}>\psi (v_{1}).
\end{equation*}
This means that the value $\tilde{v}_{1}\equiv\inf \{v_{1}\in \lbrack
0,1]:q_{1}^{\ast }(v_{1})>0\}<v^*_1$.

Finally, we show that if $q_{1}^{\ast }(v_{1})=0$ for some $v_{1}$, then $q_{1}^{\ast }(v_{1}^{\prime })=0$ for any $v_{1}^{\prime }<v_{1}$. We have
shown that if $q_{1}^{\ast }(v_{1})=0$, then $v_{1}< v_{1}^{\ast }$, and thus 
\begin{equation*}
\Pi (0,v_{1})=\psi (v_{1})+\int_{-\infty }^{-\psi (v_{1})}F(v_{2}|0)\rmd 
v_{2}=\psi (v_{1})-\psi (v_{1})=0.  
\end{equation*}
In addition, $q_{1}^{\ast }(v_{1})=0$ implies that $\Pi (q_{1},v_{1})\leq
\Pi (0,v_{1})=0$ for all $q_1\in (0,1)$. When $v_{1}^{\prime }<v_{1}$, $\psi (v_{1}^{\prime })<\psi
(v_{1})<0$, and thus for any $q_{1}\in (0,1)$, 
we have 
\begin{eqnarray*}
\Pi (q_{1},v_{1}^{\prime }) &=&\psi (v_{1}^{\prime })+(1-q_{1})\int_{-\infty
}^{-\psi (v_{1}^{\prime })}F(v_{2}|q_{1})\rmd v_{2} \\
&=&\psi (v_{1}^{\prime })+(1-q_{1})\left[ \int_{-\infty }^{-\psi
(v_{1})}F(v_{2}|q_{1})\rmd v_{2}+\int_{-\psi (v_{1})}^{-\psi (v_{1}^{\prime
})}\underset{<1}{\underbrace{F(v_{2}|q_{1})}}\rmd v_{2}\right] \\
&<&\psi (v_{1}^{\prime })+(1-q_{1})\left[ \int_{-\infty }^{-\psi
(v_{1})}F(v_{2}|q_{1})\rmd v_{2}+\psi (v_{1})-\psi (v_{1}^{\prime })\right]
\\
&<&\psi (v_{1}^{\prime })+(1-q_{1})\int_{-\infty }^{-\psi
(v_{1})}F(v_{2}|q_{1})\rmd v_{2}+\psi (v_{1})-\psi (v_{1}^{\prime }) \\
&=&\Pi (q_{1},v_{1}) \leq \Pi (0,v_{1})
=0 = \Pi (0,v_{1}^{\prime }),
\end{eqnarray*}
implying that $q_{1}^{\ast }(v_{1}^{\prime })=0$. 

As a result, for $v_1 < \tilde{v}_1$, $q^*_1(v_1) = 0$; for $v_1 \ge \tilde{v}_1$, $q^*_1(v_1)\in [0,1)$ and satisfies the first-order condition stated in the lemma; for $v_1  > \tilde{v}_1$, $q^*_1(v_1)\in (0,1)$.

To establish (ii), notice that
for $v_1<\tilde{v}_1$, $F(\cdot|q_{1}^{\ast }(v_{1}))$ reduces to a mass at $v_{2}=0$. In this case, $v_{2}=0\geq -\psi (v_{1})$ is impossible, as $-\psi (v_{1})>-\psi (\tilde{v}_{1})>-\psi (v^*_{1})=0$. Therefore, $q_{2}^{\ast}(v_{1},0)=0$. The remainder of (ii) has been established in the text. \ $\square $

\medskip


\subsection{Proof of Lemma~\ref{q-monotone}}


Fix any $v_{1} > \tilde{v}_{1}$. The problem is to choose $q_1\in [0,1]$ to maximize  
\begin{equation}
\Pi (q_{1},v_{1})=\psi
(v_{1})+(1-q_{1})\int_{-\infty }^{-\psi (v_{1})}F(v_{2}|q_{1})\rmd v_{2}.
\label{ape2}
\end{equation}
Note that restricting the range of $q_{1}$ to $(0,1)$ is without loss of generality, because (i) by the definition of $\tilde{v}_{1}$, $q_{1}=0$ cannot be optimal; and (ii) by Lemma \ref{lm-zerocutoff}, $q_{1}=1$ is not optimal either.

Define a function $\xi:
\mathbb{R} \times (0,1)\rightarrow \mathbb{R}$ as 
\begin{align}\label{ape2'}
\xi (v_{2},q_{1}) & = M(v_{2},q_{1}) \cdot (1-q_{1}) \cdot f(v_{2}|q_{1}) 
\notag \\
& = -F(v_{2}|q_{1})+(1-q_{1})\frac{\partial F(v_{2}|q_{1})}{ \partial q_{1}}.
\end{align}
Since the maximizer $q_{1}^{\ast }(v_{1})\in (0,1)$, it satisfies the first-order
condition of (\ref{ape2}) with respect to $q_{1}$:
\begin{equation}
\frac{\partial \Pi(q_{1},v_{1})}{\partial q_{1}}|_{q_{1}=q_{1}^{\ast
}(v_{1})}=\int_{-\infty }^{-\psi (v_{1})}\xi (v_{2},q_{1}^{\ast }(v_{1}))
\rmd v_{2}=0\text{.}  \label{ape3}
\end{equation}
By the second-order condition, $\frac{\partial^2 \Pi (q_{1},v_{1})}{\partial
q_{1}^2} |_{q_{1}=q_{1}^{\ast }(v_{1})} \leq 0$. 

Note that 
\begin{equation}\label{eq:xi}
\xi (-\psi (v_{1}),q_{1}^{\ast }(v_{1})) < 0,
\end{equation}
as otherwise $\xi
(v_{2},q_{1}^{\ast }(v_{1})) > 0$ for any $v_{2}<-\psi (v_{1})$ by
Assumption \ref{as:singlecrossing}, which violates (\ref{ape3}). It implies
that 
\begin{equation}
\frac{\partial^2 \Pi (q_{1},v_{1})}{\partial q_{1} \partial v_{1}}
|_{q_{1}=q_{1}^{\ast}(v_{1})} = -\psi ^{\prime }(v_{1})\xi (-\psi
(v_{1}),q_{1}^{\ast }(v_{1}))>0.  \label{ape4}
\end{equation}
Now differentiating the first-order condition (\ref{ape3}) with respect to $%
v_{1}$ on both sides of the equation leads to 
\begin{equation*}
\frac{\rmd q_{1}^{\ast }(v_{1})}{\rmd v_{1}}\cdot \underset{\leq 0}{ 
\underbrace{ \frac{\partial^2 \Pi (q_{1},v_{1})}{\partial q_{1}^2}
|_{q_{1}=q_{1}^{\ast }(v_{1})} } } + \underset{>0}{\underbrace{\frac{
\partial^2 \Pi (q_{1},v_{1})}{\partial q_{1} \partial v_{1}}
|_{q_{1}=q_{1}^{\ast}(v_{1})} }} = 0,
\end{equation*}%
which further implies that $\frac{\rmd q_{1}^{\ast }(v_{1})}{\rmd v_{1}}>0$.
\ \ $\square$

\medskip


\subsection{Proof of the claim in Remark \ref{ft1}}


Pick any $v_{1},v_{1}^{\prime }\in [\tilde{v}_1,1]$ with $v_{1}<v_{1}^{\prime
} $. Let $q_{1}^{\ast }$ and $q_{1}^{\ast \prime }$ be a maximizer of $\Pi(q_{1},v_{1})$  and $\Pi (q_{1},v_{1}^{\prime })$, respectively. By Lemma~\ref{lm-zerocutoff}, $q_{1}^{\ast }\in [0,1)$ (since $v_1$ may be $\tilde{v}_1$) and $q_{1}^{\ast \prime }\in (0,1)$. Our goal is to show that $q_{1}^{\ast }<q_{1}^{\ast \prime }$. 
Hence, it is without loss to focus on the case that $q_{1}^{\ast }, q_{1}^{\ast \prime } \in (0,1)$. 

Note that (\ref{ape4})
still applies because it only uses Assumption \ref{as:singlecrossing}. Hence, 
\begin{equation*}
\xi (-\psi (v_{1}),q_{1}^{\ast })<0 \quad \text{ and } \quad \xi (-\psi
(v_{1}^{\prime}),q_{1}^{\ast \prime })<0. 
\end{equation*}

Suppose to the contrary that $q_{1}^{\ast }\geq q_{1}^{\ast \prime }$. We
claim that 
\begin{equation}
\xi (-\psi (s),q_{1})<0\text{ for any }s\in \lbrack v_{1},v_{1}^{\prime }]%
\text{ and any }q_{1}\in \lbrack q_{1}^{\ast \prime },q_{1}^{\ast }].
\label{ape6}
\end{equation}
To see this, recall that $F(\cdot | \cdot)$ satisfies the rotation order. If 
$-\psi (s)>0$, then $\xi (-\psi (s),q_{1})<0$ for any $q_{1}\in \lbrack
q_{1}^{\ast \prime },q_{1}^{\ast }]$.

\begin{itemize}
\item If $-\psi (v_{1}^{\prime })>0$, then $-\psi (s) > 0$ for any $s\in
\lbrack v_{1},v_{1}^{\prime }]$, so (\ref{ape6}) holds.

\item Suppose that $-\psi (v_{1}^{\prime })\leq 0$. Since $\xi
(-\psi(v_{1}^{\prime }),q_{1}^{\ast \prime })<0$, the additional condition
mentioned in Remark \ref{ft1} implies that $\xi (-\psi
(v_{1}^{\prime}),q_{1})<0$ for any $q_{1}\in \lbrack q_{1}^{\ast \prime
},q_{1}^{\ast }]$. Then due to Assumption \ref{as:singlecrossing} and the
monotonicity of $\psi$, $\xi (-\psi (s),q_{1})<0$ for any $q_{1} \in \lbrack
q_{1}^{\ast \prime },q_{1}^{\ast }]$ and $s\in \lbrack v_{1},v_{1}^{\prime
}] $. This establishes (\ref{ape6}).
\end{itemize}

Due to the definition of $q_{1}^{\ast }$, $\Pi (q_{1}^{\ast },v_{1})\geq \Pi
(q_{1}^{\ast \prime },v_{1})$. If $q_{1}^{\ast }>q_{1}^{\ast \prime }$, then 
\begin{eqnarray*}
0 &\leq &\Pi (q_{1}^{\ast },v_{1})-\Pi (q_{1}^{\ast \prime
},v_{1})=\int_{q_{1}^{\ast \prime }}^{q_{1}^{\ast }}\frac{\partial \Pi
(q_{1},v_{1})}{\partial q_{1}}\rmd q_{1} \\
&=&\int_{q_{1}^{\ast \prime }}^{q_{1}^{\ast }}\left( \int_{-\infty }^{-\psi
(v_{1})}\xi (v_{2},q_{1})\rmd v_{2}\right) \rmd q_{1} \\
&=&\int_{q_{1}^{\ast \prime }}^{q_{1}^{\ast }}\left( \int_{-\infty }^{-\psi
(v_{1}^{\prime })}\xi (v_{2},q_{1})\rmd v_{2}+\int_{-\psi (v_{1}^{\prime
})}^{-\psi (v_{1})}\underset{<0\text{ by (\ref{ape6})}}{\underbrace{\xi
(v_{2},q_{1})}}\rmd v_{2}\right) \rmd q_{1} \\
&<&\int_{q_{1}^{\ast \prime }}^{q_{1}^{\ast }}\left( \int_{-\infty }^{-\psi
(v_{1}^{\prime })}\xi (v_{2},q_{1})\rmd v_{2}\right) \rmd q_{1} \\
&=&\int_{q_{1}^{\ast \prime }}^{q_{1}^{\ast }}\frac{\partial \Pi
(q_{1},v_{1}^{\prime })}{\partial q_{1}}\rmd q_{1}=\Pi (q_{1}^{\ast
},v_{1}^{\prime })-\Pi (q_{1}^{\ast \prime },v_{1}^{\prime }).
\end{eqnarray*}%
However, $\Pi (q_{1}^{\ast },v_{1}^{\prime })-\Pi (q_{1}^{\ast \prime
},v_{1}^{\prime })>0$ contradicts the optimality of $q_{1}^{\ast \prime }$.

Finally, what is left to show is that $q_{1}^{\ast }=q_{1}^{\ast \prime }$
also leads to a contradiction. In fact, if $q_{1}^{\ast }=q_{1}^{\ast \prime
}$, the first-order condition (\ref{ape3}) implies that 
\begin{equation*}
\int_{-\infty }^{-\psi (v_{1})}\xi (v_{2},q_{1}^{\ast })\rmd v_{2}=0 =
\int_{-\infty }^{-\psi (v_{1}^{\prime })}\xi (v_{2},q_{1}^{\ast })\rmd v_{2} 
\text{.}
\end{equation*}%
However, (\ref{ape6}) implies that $\xi (v_{2},q_{1}^{\ast })<0$ for any $%
v_{2}\in \lbrack -\psi (v_{1}^{\prime }),-\psi (v_{1})]$. Then,%
\begin{eqnarray*}
0 &=&\int_{-\infty }^{-\psi (v_{1})}\xi (v_{2},q_{1}^{\ast })\rmd v_{2}=%
\underset{=0}{\underbrace{\int_{-\infty }^{-\psi (v_{1}^{\prime })}\xi
(v_{2},q_{1}^{\ast })\rmd v_{2}}}+\int_{-\psi (v_{1}^{\prime })}^{-\psi
(v_{1})}\xi (v_{2},q_{1}^{\ast })\rmd v_{2} \\
&=&\int_{-\psi (v_{1}^{\prime })}^{-\psi (v_{1})}\xi (v_{2},q_{1}^{\ast }) %
\rmd v_{2}<0,
\end{eqnarray*}%
which is a contradiction. This completes the proof of the claim in Remark %
\ref{ft1}. \ $\square $

\medskip



\subsection{Proof of Lemma \ref{lem:transfer}}

We first construct the payment rule $t^{\ast }$. By Lemmas \ref{1stIC} and \ref{lm-zerocutoff} and $U(0,0)=0$, for each $v_1\in [0,1]$, we have
\begin{align}\notag
U(v_{1},v_{1}) &=U(0,0)+\int_{0}^{v_{1}}\left[ q_{1}^{\ast
}(x)+\int_{-\infty }^{+\infty }q_{2}^{\ast }(x,v_{2})F(\rmd %
v_{2}|q_{1}^{\ast }(x))\right] \rmd x \\ \notag
& = \int_{0}^{v_{1}}\left[ q_{1}^{\ast }(x)+\int_{-\psi
(x)}^{+\infty }(1-q_{1}^{\ast }(x))F(\rmd v_{2}|q_{1}^{\ast }(x))\right] %
\rmd x  \notag \\
& =  \int_{0}^{v_{1}} \bigg[ q_{1}^{\ast }(x)+(1-q_{1}^{\ast }(x)) \left(
1-F(-\psi (x)|q_{1}^{\ast }(x)) \right) \bigg] \rmd x  \notag \\ \notag
& =  \int_{0}^{v_{1}} \bigg[ 1-(1-q_{1}^{\ast }(x))F(-\psi (x)|q_{1}^{\ast
}(x)) \bigg] \rmd x   \\  \label{ape7}
& = v_{1}-\int_{0}^{v_{1}} \big( 1-q_{1}^{\ast }(x) \big) F(-\psi
(x)|q_{1}^{\ast }(x))\rmd x.  
\end{align}

On the other hand, from the envelope condition (\ref{e2'}) in the second
stage, 
\begin{align*}
 \quad (v_{1}+v_{2})q_{2}^{\ast }(v_{1},v_{2})-t^{\ast }(v_{1},v_{2}) 
& =\tilde{\pi} (v_{1},v_{2},v_{2} ) \\
& = \tilde{\pi}(v_{1}, -\psi (v_{1}), -\psi (v_{1}) )+\int_{-\psi
(v_{1})}^{v_{2}}q_{2}^{\ast }(v_{1},s)\rmd s.
\end{align*}
Thus,
\begin{equation*}
t^{\ast }(v_{1},v_{2})=(v_{1}+v_{2})q_{2}^{\ast }(v_{1},v_{2}) - \int_{-\psi
(v_{1})}^{v_{2}}q_{2}^{\ast }(v_{1},s)\rmd s-\tilde{\pi}(v_{1}, -\psi
(v_{1}), -\psi (v_{1}) ).
\end{equation*}
%
%
%
%
%
%
%
%
%
%

\begin{itemize}
\item When $v_{2} < -\psi (v_{1})$, $q_{2}^{\ast }(v_{1},v_{2}) = 0$ and 
\begin{align*}
t^{\ast }(v_{1},v_{2})& = -\tilde{\pi}(v_{1}, -\psi(v_{1}), -\psi(v_{1}) ).
\end{align*}

\item When $v_{2} \geq -\psi (v_{1})$ and $v_{1} \ge \tilde{v}_{1}$, $%
q_{2}^{\ast }(v_{1},v_{2})=1-q_{1}^{\ast}(v_{1})$ and 
\begin{align*}
t^{\ast }(v_{1},v_{2})& =(v_{1}+v_{2})(1-q_{1}^{\ast
}(v_{1}))-\int_{-\psi(v_{1})}^{v_{2}}(1-q_{1}^{\ast }(v_{1}))\rmd s-\tilde{%
\pi}(v_{1}, -\psi(v_{1}), -\psi(v_{1}) ) \\
& =\frac{1-G(v_{1})}{g(v_{1})}(1-q_{1}^{\ast }(v_{1}))-\tilde{\pi} (v_{1},
-\psi(v_{1}), -\psi(v_{1}) ).
\end{align*}

\item When $v_{1} < \tilde{v}_{1}$, $-\psi(v_{1}) > 0$, $q_{1}^{\ast}(v_{1})=0$, and $F(\cdot |q_{1}^{\ast }(v_{1})$) reduces to a mass at $0$, implying that $v_{2} \ge -\psi (v_{1}) > 0$ is impossible. As a
result, with probability one $v_{2} < -\psi (v_{1})$, which implies that 
\begin{equation*}
q_{2}^{\ast }(v_{1},v_{2})=0 \qquad \mbox{and} \qquad t^{\ast }(v_{1},v_{2})
= - \tilde{\pi}(v_{1}, -\psi(v_{1}), -\psi(v_{1}) ).
\end{equation*}
\end{itemize}

To construct the payment rule $t^{\ast}
$, it remains to pin down $\tilde{\pi}(v_{1}, -\psi(v_{1}), -\psi(v_{1}) )$. To this
end, notice that by the definition of the first-stage expected payoff, 
\begin{eqnarray*}
U(v_{1},v_{1}) &=&q_{1}^{\ast }(v_{1})\int_{-\infty }^{+\infty
}(v_{1}+v_{2})F(\rmd v_{2}|q_{1}^{\ast }(v_{1}))+\int_{-\infty }^{+\infty } 
\tilde{\pi}(v_{1},v_{2},v_{2})F(\rmd v_{2}|q_{1}^{\ast }(v_{1})) \\
&=&q_{1}^{\ast }(v_{1})v_{1}+\int_{-\infty }^{+\infty } \bigg[ %
(v_{1}+v_{2})q_{2}^{\ast }(v_{1},v_{2})-t^{\ast }(v_{1},v_{2}) \bigg] F(\rmd %
v_{2}|q_{1}^{\ast }(v_{1})) \\
&=&q_{1}^{\ast }(v_{1})v_{1}+\int_{-\psi (v_{1})}^{+\infty }(1-q_{1}^{\ast
}(v_{1}))\left[ v_{1}+v_{2}-\frac{1-G(v_{1})}{g(v_{1})}\right] F(\rmd %
v_{2}|q_{1}^{\ast }(v_{1})) \\
&& + \tilde{\pi}(v_{1}, -\psi(v_{1}), -\psi(v_{1}) ) \\
&=&q_{1}^{\ast }(v_{1})v_{1}+\int_{-\psi (v_{1})}^{+\infty }(1-q_{1}^{\ast
}(v_{1}))\left( \psi (v_{1})+v_{2}\right) F(\rmd v_{2}|q_{1}^{\ast }(v_{1}))
\\
&& + \tilde{\pi}(v_{1}, -\psi(v_{1}), -\psi(v_{1}) ).
\end{eqnarray*}
Comparing with (\ref{ape7}), we obtain 
\begin{align*}
& - \tilde{\pi}(v_{1}, -\psi(v_{1}), -\psi(v_{1}) ) \\
=& q_{1}^{\ast} (v_{1})v_{1}+\int_{-\psi (v_{1})}^{+\infty }(1-q_{1}^{\ast
}(v_{1}))[\psi (v_{1})+v_{2}]F(\rmd v_{2}|q_{1}^{\ast }(v_{1})) \\
& -\left[ v_{1}-\int_{0}^{v_{1}}(1-q_{1}^{\ast }(x))F(-\psi (x)|q_{1}^{\ast}
(x))\rmd x\right] \\
=& \int_{-\psi (v_{1})}^{+\infty }(1-q_{1}^{\ast }(v_{1}))[\psi
(v_{1})+v_{2}]F(\rmd v_{2}|q_{1}^{\ast }(v_{1}))-v_{1}(1-q_{1}^{\ast}
(v_{1})) \\
& +\int_{0}^{v_{1}}(1-q_{1}^{\ast }(x))F(-\psi (x)|q_{1}^{\ast }(x))\rmd x \\
=& (1-q_{1}^{\ast }(v_{1}))\left\{ 
\begin{array}{c}
[1-F(-\psi (v_{1})|q_{1}^{\ast }(v_{1}))]\psi (v_{1})-v_{1} \\ 
+\int_{-\psi (v_{1})}^{+\infty }v_{2}F(\rmd v_{2}|q_{1}^{\ast }(v_{1}))%
\end{array}%
\right\} \\
& +\int_{{0}}^{v_{1}}(1-q_{1}^{\ast }(x))F(-\psi (x)|q_{1}^{\ast }(x))\rmd x
\\
\overset{E[v_{2}]=0}{=}& (1-q_{1}^{\ast }(v_{1}))\left\{ 
\begin{array}{c}
\lbrack 1-F(-\psi (v_{1})|q_{1}^{\ast }(v_{1}))]\psi (v_{1})-v_{1} \\ 
-\int_{-\infty }^{-\psi (v_{1})}v_{2}F(\rmd v_{2}|q_{1}^{\ast }(v_{1}))%
\end{array}%
\right\} \\
& +\int_{{0}}^{v_{1}}(1-q_{1}^{\ast }(x))F(-\psi (x)|q_{1}^{\ast }(x))\rmd x
\\
\overset{Lemma~\ref{lem-parts}}{=}& (1-q_{1}^{\ast }(v_{1}))\left\{ 
\begin{array}{c}
\lbrack 1-F(-\psi (v_{1})|q_{1}^{\ast }(v_{1}))]\psi (v_{1})-v_{1} \\ 
+\psi (v_{1})F(-\psi (v_{1})|q_{1}^{\ast }(v_{1}))+\int_{-\infty }^{-\psi
(v_{1})}F(v_{2}|q_{1}^{\ast }(v_{1}))\rmd v_{2}%
\end{array}%
\right\} \\
& +\int_{0}^{v_{1}}(1-q_{1}^{\ast }(x))F(-\psi (x)|q_{1}^{\ast }(x))\rmd x \\
= & (1-q_{1}^{\ast }(v_{1})) \bigg[ \int_{-\infty }^{-\psi
(v_{1})}F(v_{2}|q_{1}^{\ast }(v_{1}))\rmd v_{2}-\frac{1-G(v_{1})}{g(v_{1})} %
\bigg] \\
& +\int_{0}^{v_{1}}(1-q_{1}^{\ast }(x))F(-\psi (x)|q_{1}^{\ast }(x))\rmd x.
\end{align*}
Thus, 
\begin{equation*}
t^{\ast }(v_{1},v_{2}) = 
\begin{cases}
(1-q_{1}^{\ast }(v_{1})) p^*_2(v_1) +p_{1}^{\ast }(v_{1}),
& \mbox{ if } \psi (v_{1}) + v_{2}\geq 0, \\ 
p_{1}^{\ast }(v_{1}), & \text{ otherwise},
\end{cases}
\end{equation*}
where
\begin{align*}
p_{1}^{\ast }(v_{1}) & =-\tilde{\pi}(v_{1},-\psi(v_1), -\psi(v_1))
=(1-q_{1}^{\ast
}(v_{1}))[\int_{-\infty }^{-\psi (v_{1})}F(v_{2}|q_{1}^{\ast }(v_{1}))\rmd %
v_{2}-\frac{1-G(v_{1})}{g(v_{1})}] \\
& \quad +\int_{0}^{v_{1}}(1-q_{1}^{\ast }(x))F(-\psi (x)|q_{1}^{\ast }(x))
\rmd x
\end{align*}
and $p^*_2(v_1)=\frac{1-G(v_{1})}{g(v_{1})}$. \ $\square $

\subsection{Proof of Proposition \ref{prop1}}

We need to show that the buyer has the incentive to follow the ``recommendation" that for each $v_1\in [0,1]$, (i) if type-$v_1$ buyer chooses the option contract $\{p^*_1 (v_1),q^*_1(v_1);p^*_2(v_1)\}$, he should buy the remaining $1-q^*_1(v_1)$ portion in the second stage if and only if $v_1+v_2\ge p^*_2(v_1)$; (ii) type-$v_1$ buyer should find it optimal to choose the option contract $\{p^*_1 (v_1),q^*_1(v_1);p^*_2(v_1)\}$. The verification of (i) is trivial. Thus, we only need to establish (ii) in this proof. 

As we have defined in the text, 
\begin{equation*}
w(q_1, p_2,v_1) = q_1 v_1 + \int ^{+\infty}_{p_2-v_1} (v_1 + v_2 - p_2) (1-q_1) F(\rmd v_2 | q_1).
\end{equation*}
If type-$v_1$ buyer chooses the contract $\{p^*_1 (r_1),q^*_1(r_1);p^*_2(r_1)\}$ for some $r_1$, he will buy the remaining $1-q_{1}^{\ast }(r_{1})$ portion in the second stage if and only if $v_{1}+v_{2}\geq p^*_2(r_1)$; that is, $v_{2}\geq -v_{1}+p^*_2(r_1)$. Hence, his expected utility when selecting $\{p^*_1 (r_1),q^*_1(r_1);p^*_2(r_1)\}$ and following the optimal second-stage strategy is given by
\begin{equation*}
U(v_1,r_1)=w(q^*_1(r_1), p^*_2(r_1),v_1) -p^*_1(r_1).
\end{equation*} Our goal is to show that $\Delta(v_1, r_1) \equiv U(v_1,v_1)-U(v_1,r_1)\geq 0$, for any $v_1,r_1\in [0,1]$.

To this end, notice that 
\begin{equation}\label{eq:mainprop4}
w_{3}(q_{1},p_{2},v_1)=1-(1-q_{1})F(p_{2}-v_{1}|q_{1})\ge 0.
\end{equation}
By the construction of option contracts, it is easy to verify that when the type-$v_1$ buyer chooses the contract $\{p^*_1 (v_1),q^*_1(v_1);p^*_2(v_1)\}$, his expected utility $U(v_1,v_1)$ can be expressed as the form in (\ref{ape7}). Therefore, we have
\begin{align}\label{eq:mainprop1.5}
\frac{\rmd U(v_1,v_1)}{\rmd v_1}=1-(1-q^*_1(v_1))F(-\psi(v_1)|q^*_1(v_1))
&\overset{(\ref{eq:mainprop4})}{=}w_{3}(q^*_{1}(v_1),p^*_{2}(v_1),v_1).
\end{align}

For any $v_1$ and $r_1\in [0,1]$, $\Delta(v_1, r_1) \equiv U(v_1,v_1)-U(v_1,r_1)$ is further equal to
\begin{align*}
  & U(v_1,v_{1})-U(r_1,r_1) + U(r_1,r_1) -U(v_1,r_{1}) \\
= & U(v_1,v_{1})-U(r_1,r_1) + [ w(q^*_1(r_1), p^*_2(r_1),r_1) -p^*_1(r_1) ] -[ w(q^*_1(r_1), p^*_2(r_1),v_1) -p^*_1(r_1) ] \\
\overset{(\ref{eq:mainprop1.5})} {=} & \int^{v_1}_{r_1}w_3(q^*_1(s), p^*_2(s),s) \rmd s 
+  w(q^*_1(r_1), p^*_2(r_1),r_1) 
- w(q^*_1(r_1), p^*_2(r_1),v_1) \\
= & \int^{v_1}_{r_1} [w_3(q^*_1(s), p^*_2(s),s) -  w_3(q^*_1(r_1), p^*_2(r_1),s) ]\rmd s \\
= & \int^{v_1}_{r_1} \int^{s}_{r_1} 
\left[ 
w_{31}(q^*_1(x),  p^*_2(x),s){q_1^*}'(x) 
+w_{32}(q^*_1(x), p^*_2(x),s){p_2^*}'(x) 
\right] \rmd x \rmd s   \\
= & \int^{v_1}_{r_1} \int^{v_1}_{x} 
\left[ 
w_{31}(q^*_1(x), p^*_2(x),s){q_1^*}'(x) 
+w_{32}(q^*_1(x), p^*_2(x),s){p_2^*}'(x) 
\right] \rmd s  \rmd x, 
\end{align*}
where the third and the fourth equalities follow from the fundamental theorem of calculus and the last equality interchanges the order of integration. 

Since ${p_2^*}'(x)\le 0$ and $w_{32}(q^*_1(x), p^*_2(x),s)\le 0$ for all $x$, $s\in [0,1]$, to establish $\Delta (v_{1},r_{1})  \ge 0$, it suffices to show that 
\begin{equation*}
\int^{v_1}_{r_1} \int^{v_1}_{x}  w_{31}(q^*_1(x), p^*_2(x),s) {q_1^*}'(x)\rmd s \rmd x \ge 0.
\end{equation*}
Notice that
\begin{align*}
\int^{v_1}_{x}  w_{31}(q^*_1(x), p^*_2(x),s)\rmd s 
& \overset{(\ref{eq:mainprop4})}{=} \int^{v_1}_{x}  \frac{\partial [1- (1-q^*_1(x)) F(p^*_2(x)-s | q^*_1(x)) ]}{\partial q^*_1(x)} \rmd s \\
& \overset{(\ref{ape2'})}{=} -\int^{v_1}_{x} \xi(p^*_2(x) - s ,q^*_1(x)) \rmd s 
\\
& = \int^{x-v_1-\psi(x)}_{-\psi(x)} \xi( y ,q^*_1(x)) \rmd y, 
\end{align*}
where the last equality follows the definition of $p^*_2(x)$ and the change of variable $y=\frac{1-G(x)}{g(x)} - s=x-s-\psi(x)$. 
Hence, to show $\Delta (v_{1},r_{1})  \ge 0$, it suffices to show
\begin{equation}\label{eq:mainprop5}
\int^{v_1}_{r_1} {q_1^*}'(x) \int^{x-v_1-\psi(x)}_{-\psi(x)} \xi( y ,q^*_1(x)) \rmd y  \rmd x \ge 0.
\end{equation}

Now we discuss two cases and show that in both cases expression (\ref{eq:mainprop5}) holds; therefore $\Delta (v_{1},r_{1})\geq 0$.

Case 1: for $r_1 \ge v_1$ and $x \in [v_1, r_1]$, we must have $-\psi (x) \le  x-v_{1}-\psi (x)$.

If $x$ is such that $q_{1}^{\ast }(x)=0$, it has been established that ${q_1^*}'(x)=0$. Thus,
$$
{q_1^*}'(x) \int^{x-v_{1}-\psi (x)}_{-\psi
	(x)}\xi (y,q_{1}^{\ast}(x))\rmd y =  0.$$ 

If $x$ is such that $q_{1}^{\ast }(x)>0$, by (\ref{ape4}), $\xi (-\psi (x),q_{1}^{\ast }(x))<0$. Assumption \ref{as:singlecrossing}
implies that $\xi (y,q_{1}^{\ast }(x))<0$ for any $y\geq -\psi (x)$. Also, notice that ${q_1^*}'(x)\ge 0$.
Thus,
$$
{q_1^*}'(x) \int^{x-v_{1}-\psi (x)}_{-\psi
	(x)}\xi (y,q_{1}^{\ast}(x))\rmd y \le  0.$$

Since $r_1 \ge v_1$, expression (\ref{eq:mainprop5}) holds.

Case 2: for $r_1 \le v_1$ and $x \in [ r_1, v_1]$, we have $-\psi (x) \ge  x-v_{1}-\psi (x)$.

If $x$ is such that $q_{1}^{\ast }(x)=0$, again,
$$
{q_1^*}'(x) \int^{x-v_{1}-\psi (x)}_{-\psi
	(x)}\xi (y,q_{1}^{\ast}(x))\rmd y =  0.$$

If $x$ is such that $q_{1}^{\ast }(x)>0$ and $\xi (x-v_{1}-\psi (x),q_{1}^{\ast }(x))\leq 0$, then Assumption \ref{as:singlecrossing} implies that $\xi (y,q_{1}^{\ast }(x)) \le 
	0$ for any $y \ge x-v_{1}-\psi (x)$, which further implies that 
$$
{q_1^*}'(x) \int^{x-v_{1}-\psi (x)}_{-\psi
	(x)}\xi (y,q_{1}^{\ast}(x))\rmd y \ge  0.$$ 
	
If $x$ is such that $q_{1}^{\ast }(x)>0$ and $\xi (x-v_{1}-\psi (x),q_{1}^{\ast }(x))>0$, then Assumption \ref{as:singlecrossing} implies that $\xi (y,q_{1}^{\ast }(x)) \ge 0$
	for any $y\le x-v_{1}-\psi (x)$. Since $q_{1}^{\ast }(x)>0$, from (\ref{ape3}), the optimality of $q_{1}^{\ast }(x)$ requires $
	\int_{-\infty }^{-\psi (x)}\xi (y,q_{1}^{\ast }(x))\rmd y=0$. It
	then follows that 
	\begin{equation*}
	0=\int_{-\infty }^{-\psi (x)}\xi (y,q_{1}^{\ast }(x))\rmd y=\underset%
	{\ge 0}{\underbrace{\int_{-\infty }^{x-v_{1}-\psi (x)}\xi
			(y,q_{1}^{\ast }(x))\rmd y}}+\int_{x-v_{1}-\psi (x)}^{-\psi
		(x)}\xi (y,q_{1}^{\ast }(x))\rmd y,
	\end{equation*}
	which implies that $\int_{x-v_{1}-\psi (x)}^{-\psi (x)}\xi
	(y,q_{1}^{\ast }(x))\rmd y\le 0$, \textit{i.e.}, 
$$
{q_1^*}'(x) \int^{x-v_{1}-\psi (x)}_{-\psi
	(x)}\xi (y,q_{1}^{\ast}(x))\rmd y \ge  0.$$

Expression (\ref{eq:mainprop5}) holds again, since $r_1\le v_1$.

In both Cases 1 and 2, we conclude that $\Delta (v_{1},r_{1})\geq 0$. This completes the proof.  \ $\square $

\subsection{Proof of Lemma \ref{lem:payment}}


For (i), notice that for $v_1<\tilde{v}_1$, $q_{1}^{\ast }(v_{1})=0$. In this case, $F(\cdot |0)$ degenerates to a mass at $0$ and with
probability one $\psi (v_{1})+v_{2} = \psi (v_{1}) < 0$, where the inequality follows from Lemma~\ref{lm-zerocutoff}. As a result,
\begin{eqnarray}
p_{1}^{\ast }(v_{1}) &=&\int_{-\infty }^{-\psi (v_{1})}F(v_{2}|0)\rmd v_{2}-%
\frac{1-G(v_{1})}{g(v_{1})}+\int_{0}^{v_{1}}F(-\psi (x)|0)\rmd x  \notag \\
&=& -\psi (v_{1})-\frac{1-G(v_{1})}{g(v_{1})}+v_{1}  \notag \\
&=& 0.  \label{0payment}
\end{eqnarray}
 By Lemma \ref{lm-zerocutoff}, there is no consumption in both
stages when $v_{1}\in [0,\tilde{v}_{1})$. Then by Lemma \ref{lem:transfer}, when $v_{1} < \tilde{v}_{1}$, 
\begin{equation*}
t^{\ast }(v_{1},0)=p_{1}^{\ast }(v_{1})=0\text{.}  
\end{equation*}


For (ii), when $v_{1}\in [\tilde{v}_1,1]$,
\begin{align*}
p_{1}^{\ast \prime }(v_{1}) = & -\psi ^{\prime }(v_{1})(1-q_{1}^{\ast
}(v_{1}))F(-\psi (v_{1})|q_{1}^{\ast }(v_{1})) \\
& +q_{1}^{\ast \prime }(v_{1}) \underset{=0\text{ by (\ref{ape3})}}{%
	\underbrace{\int_{-\infty }^{-\psi (v_{1})}\frac{\partial (1-q_{1}^{\ast
			}(v_{1}))F(v_{2}|q_{1}^{\ast }(v_{1})) }{\partial q_{1}^{\ast }(v_{1})}\rmd %
		v_{2}}} \\
& -(1-q_{1}^{\ast }(v_{1}))(\frac{1-G(v_{1})}{g(v_{1})})^{\prime
}+q_{1}^{\ast \prime }(v_{1})\frac{1-G(v_{1})}{g(v_{1})} \\
& +(1-q_{1}^{\ast }(v_{1}))F(-\psi (v_{1})|q_{1}^{\ast }(v_{1})) \\
= & q_{1}^{\ast \prime }(v_{1})\frac{1-G(v_{1})}{g(v_{1})}-(1-q_{1}^{\ast
}(v_{1}))(1-F(-\psi (v_{1})|q_{1}^{\ast }(v_{1})))(\frac{1-G(v_{1})}{g(v_{1})%
})^{\prime }.
\end{align*}
Recall that for $v_{1}\in [\tilde{v}_1,1]$, $q_{1}^{\ast \prime }(v_{1})>0$, $\frac{1-G(v_{1})}{g(v_{1})}\ge 0$ with strict inequality when $v_1\in [\tilde{v}_1,1)$, and $(\frac{1-G(v_{1})}{g(v_{1})})^{\prime }<0$ (Assumption \ref{as:monotonehazardrate}). 
It can be seen that $ p_{1}^{\ast \prime }(v_{1}) \ge  0$ with strict inequality when $v_1\in [\tilde{v}_1,1)$. Hence, $p_{1}^{\ast}$ is strictly increasing on $[\tilde{v}_1,1]$.

It is easy to see that $\Pi(0,\tilde{v}_1)=\Pi(q^*_1(\tilde{v}_1),\tilde{v}_1)$. As a result,
$$\int^{-\psi(\tilde{v}_1)}_{-\infty}F(x|0)\rmd x=(1-q^*_1(\tilde{v}_1)) \int^{-\psi(\tilde{v}_1)}_{-\infty}F(x|q^*_1(\tilde{v}_1))\rmd x,$$ which, jointly with Lemma \ref{lem:transfer}, implies that 
\begin{align*}
p^*_1(\tilde{v}_1)= &\int^{-\psi(\tilde{v}_1)}_{-\infty}F(x|0)\rmd x-(1-q^*_1(\tilde{v}_1))\frac{1-G(\tilde{v}_{1})}{g(\tilde{v}_{1})}+\int_{0}^{\tilde{v}_{1}}(1-q_{1}^{\ast }(x))F(-\psi (x)|q_{1}^{\ast }(x))\rmd x \\ 
=& -\psi(\tilde{v}_1)- (1-q^*_1(\tilde{v}_1))\frac{1-G(\tilde{v}_{1})}{g(\tilde{v}_{1})} + \tilde{v}_1=q^*_1(\tilde{v}_1)\frac{1-G(\tilde{v}_1)}{g(\tilde{v}_1)}.
\end{align*}

For (iii), the result follows directly from Assumption \ref{as:monotonehazardrate}. 

For (iv), when $v_1<\tilde{v}_1$, $q_{1}^{\ast }(v_{1})=0$, 
\begin{equation*}
\underset{=0\text{ by (\ref{0payment})}}{\underbrace{p_{1}^{\ast }(v_{1})}}%
+p_{2}^{\ast }(v_{1})(1-q_{1}^{\ast }(v_{1}))=p_{2}^{\ast }(v_{1})=\frac{
	1-G(v_{1})}{g(v_{1})},
\end{equation*}
which is strictly decreasing in $v_{1}$. 

Plugging in the expressions of $p^*_1$ and $p^*_2$, we have that
\begin{align*}
& p_{1}^{\ast }(v_{1})+p_{2}^{\ast }(v_{1})(1-q_{1}^{\ast }(v_{1})) \\
=& (1-q_{1}^{\ast }(v_{1}))\int_{-\infty }^{-\psi
	(v_{1})}F(v_{2}|q_{1}^{\ast }(v_{1}))\rmd v_{2}+\int_{0}^{v_{1}}(1-q_{1}^{
	\ast }(x))F(-\psi (x)|q_{1}^{\ast }(x))\rmd x.
\end{align*}
For $v_1\ge \tilde{v}_1$, since (\ref{foc}) applies, the derivative of the above expression with
respect to $v_{1}$ is 
\begin{equation*}
(1-q_{1}^{\ast }(v_{1}))F(-\psi (v_{1})|q_{1}^{\ast }(v_{1}))\cdot (\frac{
	1-G(v_{1})}{g(v_{1})})^{\prime }<0.
\end{equation*}

For (v), the expected payment of any type $v_1$ is given by
\begin{align*}
& p_{1}^{\ast }(v_{1})+p_{2}^{\ast }(v_{1})(1-q_{1}^{\ast }(v_{1}))(1-F(-\psi(v_1)|q^*_1(v_1))) \\
=&  p_{1}^{\ast }(v_{1})+p_{2}^{\ast }(v_{1})(1-q_{1}^{\ast }(v_{1}))-p_{2}^{\ast }(v_{1})(1-q_{1}^{\ast }(v_{1}))F(-\psi(v_1)|q^*_1(v_1)).
\end{align*}
For $v_1\in [0,\tilde{v}_1)$, $p^*_1(v_1)=0$ and $F(-\psi(v_1)|q^*_1(v_1))=F(-\psi(v_1)|0)=1$ (since $\tilde{v}_1<v^*_1$). Thus, the expected payment of $v_1\in [0,\tilde{v}_1)$ is equal to zero.
For $v_1\in [\tilde{v}_1,1]$, the derivative of the above expression with respect to $v_1$ is 
\begin{align*}
&(1-q_{1}^{\ast }(v_{1}))F(-\psi (v_{1})|q_{1}^{\ast }(v_{1}))\cdot (\frac{
	1-G(v_{1})}{g(v_{1})})^{\prime }-\big(\frac{1-G(v_1)}{g(v_1)}(1-q_{1}^{\ast }(v_{1}))F(-\psi(v_1)|q^*_1(v_1))\big)'\\
=&-\frac{1-G(v_1)}{g(v_1)}  \big((1-q_{1}^{\ast }(v_{1}))F(-\psi(v_1)|q^*_1(v_1))\big)'\\
=&-\frac{1-G(v_1)}{g(v_1)}  \big( 
\underset{<0}{\underbrace{-\psi'(v_1) }}
(1-q_{1}^{\ast }(v_{1}))f(-\psi(v_1)|q^*_1(v_1)) + 
\underset{<0\text{ by (\ref{eq:xi})}}{\underbrace{\xi(-\psi(v_1),q^*_1(v_1))}{q^*_1}'(v_1)}\big)
>0.
\end{align*}

The proof completes. \ $\square $


\subsection{Proof of the claim in Section~\ref{subsec-illustrative}}


\textbf{Rotation order}. Note that 
\begin{equation*}
\frac{\partial F(v_{2}|q_{1})}{\partial q_1} = \frac{\partial
\int_{-\infty}^{\frac{v_{2}}{q_{1}}} \frac{1}{\sqrt{2 \pi}} e^{- \frac{s^2}{2%
}} \rmd s}{\partial q_1} = - \frac{v_2}{q_1^2} \frac{1}{\sqrt{2 \pi}} e^{- 
\frac{v_2^2}{2q_1^2}}.
\end{equation*}
It is clear that $F$ satisfies the rotation order, as 
\begin{equation*}
\frac{\partial F(v_{2}|q_{1})}{\partial q_{1}} 
\begin{cases}
> 0, & \text{ when } v_{2} < 0; \\ 
= 0, & \text{ when } v_{2} = 0; \\ 
< 0, & \text{ when } v_{2} > 0.%
\end{cases}%
\end{equation*}

\textbf{Assumption~\ref{as:singlecrossing}}. By Remark~\ref{rmk-crossing}, it
suffices to show that $F(v_{2}|q_{1})/f(v_{2}|q_{1})$ is increasing in $%
v_{2} $ and $\frac{\partial F(v_{2}|q_{1})}{\partial q_{1}}/f(v_{2}|q_{1})$
is decreasing in $v_{2}$. The latter is straightforward as 
\begin{equation*}
\frac{\partial F(v_{2}|q_{1})}{\partial q_{1}}/f(v_{2}|q_{1}) = \left( - 
\frac{v_2}{q_1^2} \frac{1}{\sqrt{2 \pi}} e^{- \frac{v_2^2}{2q_1^2}}
\right)/\left( \frac{1}{q_{1} \sqrt{2 \pi}} e^{- \frac{v_{2}^2}{2 q^2_{1}}}
\right) = - \frac{v_2}{q_1}.
\end{equation*}
To show that $F(v_{2}|q_{1})/f(v_{2}|q_{1})$ is increasing in $v_{2}$, note
that 
\begin{equation*}
F(v_{2}|q_{1})/f(v_{2}|q_{1}) = \left(\int_{-\infty}^{\frac{v_{2}}{q_{1}}} 
\frac{1}{\sqrt{2 \pi}} e^{- \frac{s^2}{2}} \rmd s \right)/\left(\frac{1}{%
q_{1} \sqrt{2 \pi}} e^{- \frac{v_{2}^2}{2 q^2_{1}}} \right).
\end{equation*}
By changing variables as $x = \frac{v_{2}}{q_{1}}$, one needs to show that $%
\varphi(x) = \left(\int_{-\infty}^{x} e^{- \frac{s^2}{2}} \rmd s \right) e^{%
\frac{x^2}{2}}$ is increasing in $x$. We have that 
\begin{equation*}
\varphi^{\prime }(x) = 1 + \left(\int_{-\infty}^{x} e^{- \frac{s^2}{2}} \rmd %
s \right) e^{\frac{x^2}{2}} x = e^{\frac{x^2}{2}} \left(e^{-\frac{x^2}{2}} +
x \int_{-\infty}^{x} e^{- \frac{s^2}{2}} \rmd s \right).
\end{equation*}
When $x \to -\infty$, $e^{-\frac{x^2}{2}} \to 0$, and $x \int_{-\infty}^{x}
e^{- \frac{s^2}{2}} \rmd s \to 0$ by L'H\^ospital's rule. In addition, 
\begin{equation*}
\left(e^{-\frac{x^2}{2}} + x \int_{-\infty}^{x} e^{- \frac{s^2}{2}} \rmd s
\right)^{\prime }= \int_{-\infty}^{x} e^{- \frac{s^2}{2}} \rmd s > 0.
\end{equation*}
Thus, $\varphi^{\prime }(x) \ge 0$ and $\varphi(x)$ is increasing, which
implies that Assumption~\ref{as:singlecrossing} holds.

\textbf{Assumption~\ref{as:monotonehazardrate}}. It is clear that $\frac{%
1-G(v_{1})}{g(v_{1})} = 1 - v_1$ is decreasing, implying that Assumption~\ref%
{as:monotonehazardrate} holds.

\textbf{Monotonicity $\nRightarrow$ global IC}. Finally, we show that $(\hat{%
q}_1$, $\hat{q}_2)$ cannot be the allocation rule in an incentive-compatible
mechanism. Suppose that the claim does not hold. Then there exists some $%
\hat{t}$ such that $(\hat{q}_1$, $\hat{q}_2, \hat{t})$ is IC in both stages.
We abuse the notation a bit by still using $U(v_1, r_1)$ to denote the
buyer's utility with the first-stage type $v_1$ and report $r_1$.

By Lemma~\ref{1stIC}, for $v_1 \ge \frac{1}{2}$, 
\begin{align*}
U(v_{1},v_{1}) & = \int_{0}^{v_{1}} \left[ \hat{q}_{1}(s) + \int_{-\infty
}^{+\infty } \hat{q}_{2}(s,v_{2}) F(\rmd v_{2} | \hat{q}_{1}(s)) \right] %
\rmd s \\
& = \int_{0}^{v_{1}} \left[ \hat{q}_{1}(s) + \int_{-\psi(s) }^{+\infty } (1
- \hat{q}_{1}(s)) F(\rmd v_{2} | \hat{q}_{1}(s)) \right] \rmd s \\
& = \int_{0}^{v_{1}} \bigg[ \hat{q}_{1}(s) + (1 - \hat{q}_{1}(s)) (1 -
F(-\psi(s) | \hat{q}_{1}(s)) \bigg] \rmd s \\
& = \int_{0}^{v_{1}} \bigg[ 1 - H \left( - \frac{1}{\psi(s)} \right) (1 -
\psi^2(s)) \bigg] \rmd s,
\end{align*}
where the last equality holds since $F(v_1 | q_1) = H(\frac{v_1}{q_1})$. On
the other hand, 
\begin{equation*}
U(v_{1},v_{1}) = \hat{q}_{1}(v_{1}) v_{1} + \int_{-\psi(v_{1}) }^{+\infty }
(1 - \hat{q}_{1}(v_{1}))(v_1 + v_2) F(\rmd v_{2} | \hat{q}_{1}(v_{1})) - 
\hat{T}(v_1),
\end{equation*}
where 
\begin{equation*}
\hat{T}(v_1) = \int_{- \infty }^{+\infty } \hat{t}(v_{1}, v_{2}) F(\rmd %
v_{2} | \hat{q}_{1}(v_{1})).
\end{equation*}
Then we have 
\begin{align*}
U(v_{1},r_{1}) & = \hat{q}_{1}(r_{1}) v_{1} + \int_{-\psi(r_{1}) }^{+\infty
} (1 - \hat{q}_{1}(r_{1}))(v_1 + v_2) F(\rmd v_{2} | \hat{q}_{1}(r_{1})) - 
\hat{T}(r_1) \\
& = \hat{q}_{1}(r_{1}) v_{1} + \int_{-\psi(r_{1}) }^{+\infty } (1 - \hat{q}%
_{1}(r_{1}))(v_1 + v_2) F(\rmd v_{2} | \hat{q}_{1}(r_{1})) \\
& \quad + U(r_{1},r_{1}) - \hat{q}_{1}(r_{1}) r_{1} - \int_{-\psi(r_{1})
}^{+\infty } (1 - \hat{q}_{1}(r_{1}))(r_1 + v_2) F(\rmd v_{2} | \hat{q}%
_{1}(r_{1})) \\
& = U(r_{1},r_{1}) + \hat{q}_{1}(r_{1}) (v_{1} - r_{1}) + (1 - \hat{q}%
_{1}(r_{1})) (v_{1} - r_{1}) \left( 1 - H \left( - \frac{1}{\psi(r_{1})}
\right) \right) \\
& = U(r_{1},r_{1}) + (v_{1} - r_{1}) - (1 - \psi^2(r_{1})) (v_{1} - r_{1}) H
\left( - \frac{1}{\psi(r_{1})} \right).
\end{align*}
By simple algebra, 
\begin{equation*}
U(v_{1},v_{1}) \ge U(v_{1},r_{1}) \qquad \Longleftrightarrow
\end{equation*}
\begin{equation*}
\int_{r_1}^{v_{1}} H \left( - \frac{1}{\psi(s)} \right) (1 - \psi^2(s)) \rmd %
s \le (v_{1} - r_{1}) H \left( - \frac{1}{\psi(r_{1})} \right) (1 -
\psi^2(r_{1})),
\end{equation*}
which may not be always true. We observe that $H( - \frac{1}{x} ) (1 - x^2)
\ge 0$ for $x \in [0, 1]$, converges to $0$ when either $x \to 0$ or $x \to
1 $. Thus, there must be an open set $(a_1, a_2) \subseteq [0, 1]$ such that 
$H( - \frac{1}{x} ) (1 - x^2)$ is strictly increasing on $(a_1, a_2)$. Pick $%
r_1$ and $v_1$ such that $a_1 < \psi(r_1) < \psi(v_1) < a_2$. Then for any $%
s \in (r_1, v_1]$, 
\begin{equation*}
H \left( - \frac{1}{\psi(s)} \right) (1 - \psi^2(s)) > H \left( - \frac{1}{%
\psi(r_1)} \right) (1 - \psi^2(r_1)),
\end{equation*}
which implies that $U(v_{1},v_{1}) < U(v_{1},r_{1})$. This is a contraction.
\ $\square $

\medskip


\end{document}